\documentclass{article}

\usepackage{arxiv}
\usepackage{amsmath}
\usepackage{graphicx,multicol,multirow}
\usepackage{enumerate}
\usepackage{xcolor}
\usepackage{float}
\usepackage{caption}
\usepackage{moreverb}
\usepackage{booktabs}
\usepackage{makecell}
\usepackage{tabularx}
\usepackage{varioref}
\usepackage{enumitem}
\usepackage{setspace}
\usepackage[utf8]{inputenc} 
\usepackage[T1]{fontenc}    
\usepackage{lipsum}		
\usepackage{times}
\usepackage[pagewise, left]{lineno} 
\usepackage[normalem]{ulem}  
 \usepackage{xr}
 \usepackage{ulem}
\usepackage[authoryear]{natbib}
\usepackage{url} 
\usepackage[nokeyprefix]{refstyle}
 \makeatletter
\chardef\bslash=`\\ 

\title{Information borrowing in Bayesian clinical trials: choice of tuning parameters for the robust mixture prior}
\author{
Vivienn Weru\\
	Division of Biostatistics\\
	German Cancer Research Center (DKFZ),\\
   Medical Faculty \\
   University of Heidelberg\\
	Heidelberg, Germany \\
	\texttt{vivienn.weru@dkfz-heidelberg.de} \\
	\And
	Annette Kopp-Schneider\\
    Division of Biostatistics\\
	German Cancer Research Center (DKFZ)\\
Heidelberg, Germany \\
	\And
	Manuel Wiesenfarth\\
    Cogitars GmbH \\
	Heidelberg, Germany\\
	\And
	Sebastian Weber\\
	Advanced Quantitative Sciences\\
Novartis Pharma AG \\
4002 Basel, Switzerland \\
	\And
	Silvia Calderazzo\\
    Division of Biostatistics\\
	German Cancer Research Center (DKFZ)\\
Heidelberg, Germany 
}

\begin{document}
\maketitle

\begin{abstract}
External data borrowing in clinical trial designs has increased in recent years. This is accomplished in the Bayesian framework by specifying informative prior distributions. To mitigate the impact of potential inconsistency (bias) between external and current data, robust approaches have been proposed. One such approach is the robust mixture prior arising as a mixture of an informative prior and a more dispersed prior inducing dynamic borrowing. This prior requires the choice of four quantities: the mixture weight, mean, dispersion and parametric form of the robust component. To address the challenge associated with choosing these quantities, we perform a case-by-case study of their impact on specific operating characteristics in one-arm and hybrid-control trials with a normal endpoint. All four quantities were found to strongly impact the operating characteristics. As already known, variance of the robust component is linked to robustness. Less known, however, is that its location can have severe impact on test and estimation error. Further, the impact of the weight choice is strongly linked with the robust component’s location and variance. We provide recommendations for the choice of the robust component parameters, prior weight, alternative functional form for this component and considerations for evaluating operating characteristics.
\end{abstract}


\section{Introduction}
\label{s:intro}
In clinical trials with small populations it is sometimes desired to borrow information from external sources to increase efficiency of the current study. 
Bayesian methods have seen widespread application in this context, due to the ease in incorporating external information via the specification of informative prior distributions. A concern regards borrowing when the current and the external data are different, i.e. there exists bias between true current and observed external data means, as this could lead to wrong conclusions about the current trial. \cite{pocock1976combination}, in his seminal paper, provided guidelines to ensure that the selected external data are consistent with current data. Propensity score approaches may also be used to minimize bias due to observed sources when borrowing external controls by matching to the active control patients. In particular, recent methods have been proposed to allow estimation of the average treatment effect considering potential comparability issues between the current and external data \citep{li2023improving, gao2025improving, dang2022cross, gordon2025non}. 
\cite{gao2025improving} propose a borrowing approach where external controls are weighted based on covariate information to match current trial controls. An adaptive lasso penalty is then used to implement dynamic borrowing to mitigate bias from unknown sources. 
\cite{dang2022cross} propose an approach that, using cross-validation, selects the study from a range of candidate studies, including with and without external controls, for robust estimation of average treatment effect.

Nonetheless, it is quite likely that there would still be differences in the outcomes of external and current patient populations which cannot be explicitly adjusted for. In the presence of bias between the external and current data, borrowing external data worsens operating characteristics, e.g. inflating Type I error (TIE) rate if the external data favors the alternative hypothesis. To mitigate this risk, dynamic borrowing methods have been proposed that allow borrowing most when the current and external data are observed to be similar and least otherwise. These include the power prior \citep{ibrahim2000power}, meta-analytic priors \citep{schmidli2014robust}, the robust mixture prior \citep{schmidli2014robust} and commensurate prior \citep{hobbs2011hierarchical}. \cite{calderazzo2024robust} take a different route and propose a compromise decision approach that focuses not on the prior but on test decisions and tuning these to achieve a cap on TIE rate. 

For most of these methods, the user has to specify parameters that regulate the amount of borrowing with some methods being more sensitive to observed bias than others. Nonetheless, there is no consensus on how to select a method or its parameters. In this paper we aim to investigate the impact of different parameter choices on operating characteristics, focusing on the robust mixture prior. While all the methods just mentioned allow dynamic borrowing, the robust mixture prior has been used more widely in practice perhaps facilitated by availability of user-friendly software, e.g. the RBesT \citep{weber2019applying} package in R. We will focus on trials with a normal endpoint. While standard choices for the robust component of the mixture prior are available in the binomial case (e.g. the uniform or Jeffreys prior on the response probability), the normal case lacks similarly accepted defaults. This makes the specification of the robust component more critical. 

The paper is structured as follows; an overview of model set-up and inference is provided in Section \ref{s:Bayes} while an introduction to the mixture prior is given in Section \ref{s:model}. Section \ref{s:metrics} outlines the metrics used to evaluate the borrowing approach. In Sections \ref{s:one_arm} and \ref{s:two_arm}, borrowing in the one-arm and hybrid-control trial, respectively, is presented with set-ups and results described therein. Discussion of findings and recommendations is given in Section \ref{s:discuss}. Finally, conclusions are presented in Section \ref{s:conclusion}. 
\section{Model set-up and inference}
\label{s:Bayes}
\subsection{One-arm trial}
For a one-arm trial, let
$ y=\{y_1,...,y_n\}$ be the current trial data and let $\theta$ represent the true underlying parameter of interest. Let $y_{\text{ext}}=\{y_{\text{ext},1},...,y_{\text{ext},n_{\text{ext}}}\}$ be the external data available. Throughout this paper, the external data will be considered fixed. 
With interest being inference for the parameter of interest $\theta$, we test, without loss of generality, the following one-sided hypothesis:
 \begin{gather*}
     H_0: \theta \leq \theta_0 \quad \text{versus} \quad H_1: \theta>\theta_0.
 \end{gather*}
In the Bayesian framework, inference is based on the posterior density where $\pi(\theta)$ will be considered as the prior. The decision rule is given as follows: $H_0$ is rejected if
$
    Pr^{\pi}(\theta \leq \theta_0|y)\leq\alpha
$, where $\alpha$ is often chosen such that TIE rate with a non-informative prior is same as the TIE rate of the frequentist test and $$Pr^{\pi}(\theta\leq \theta_0|y)=\int_{-\infty}^{\theta_0}\pi(\theta|y)d\theta, $$ with $\pi(\theta|y)$ being the posterior distribution of $\theta$. \cite{held2020bayesian} provides more details on the use of posterior tail probabilities for inference detailing their connection to one-sided p-values and their applicability in decision making. 
We will evaluate frequentist operating characteristics based on the Bayesian test decisions, computing TIE rate as follows:
\begin{gather}
   \int_{-\infty}^{\infty} I\{Pr^{\pi}(\theta\leq \theta_0|y)\leq\alpha\}f(y|\theta=\theta_0)dy, 
   \label{eq:TIE_onearm}
\end{gather}
where $f(y|\theta=\theta_0)$ is the probability density function of the current data under the null hypothesis and $I\{.\}$ is an indicator function. Similarly, we evaluate power at $\theta=\theta_1 \in H_1$. 

For estimation, we use the posterior mean as the point estimate $\hat{\theta}(y)$ since it is the optimal Bayes estimator under the quadratic loss \citep{robert2007bayesian}. We evaluate the performance of such estimator based on the frequentist Root Mean Square Error (RMSE) given as follows:
\begin{gather}
\sqrt{
   \int_{-\infty}^{\infty}(\hat{\theta}(y)-\theta)^2f(y|\theta) dy}.
   \label{eq:MSE}
\end{gather}
\subsection{Two-arm trial}
In the situation of a two-arm trial for which external information is available for the control arm, referred to in the literature as the hybrid-control trial (e.g. \cite{fu2023covariate, kopp2024simulating,li2022hybrid}), let $ y_c=\{y_{c,1},...,y_{c,n_c}\}$ and $ y_t=\{y_{t,1},...,y_{t,n_t}\}$
be the current control and current treatment data, respectively. Let $\theta_t$ and $\theta_c$ represent the underlying parameter of interest in the current treatment and current control groups. We test, without loss of generality, the following one-sided hypothesis:
 \begin{gather*}
     H_0: \theta_t \leq \theta_c \quad \text{versus} \quad H_1: \theta_t > \theta_c.
 \end{gather*}
The decision rule is based on the posterior distribution of the treatment effect and in particular $H_0$ is rejected if 
$
    Pr^{\pi}(\theta_t \leq \theta_c|y_t,y_c)\leq\alpha
$ where $$Pr^{\pi}(\theta_t\leq \theta_c|y_t,y_c)=\int_{-\infty}^{\theta_c}\int_{-\infty}^{\infty}\pi(\theta_c|y_c)\pi(\theta_t|y_t)d\theta_cd\theta_t. $$
TIE rate is given by:
\begin{gather}
  \int_{-\infty}^{\infty} \int_{-\infty}^{\infty}I\{Pr^{\pi}(\theta_t\leq \theta_c|y_t,y_c)\leq \alpha\} f(y_c|\theta_c=\theta_t)f(y_t|\theta_c=\theta_t)dy_cdy_t.
     \label{eq:TIE_twoarm}
\end{gather}
Similarly, we evaluate power at $\theta_t-\theta_c=\theta_1 \in H_1$.\\
One may also average TIE rate over the prior to obtain the Bayesian analogs of TIE rate and power \citep{berry2010bayesian, best2025beyond}. Average TIE rate is obtained as:
\begin{gather}
    \int_{-\infty}^{\infty}\int_{-\infty}^{\infty} \int_{-\infty}^{\infty}I\{Pr^{\pi^a}(\theta_t\leq \theta_c|y_t,y_c)\leq \alpha\} f(y_c|\theta_c=\theta_t)f(y_t|\theta_c=\theta_t) \pi^d(\theta_c)dy_cdy_t d\theta_c,
     \label{eq:avgTIE}
\end{gather}
where $\pi^d$ is the design prior used to generate the data while $\pi^a$ is the prior used at the analysis stage. 
\section{Bayesian dynamic borrowing using the mixture prior}
\label{s:model}
Robust mixture priors have emerged as an important tool to incorporate uncertainty about the degree of similarity between current and external data. Such priors have been used when incorporating external data in a current clinical trial (e.g. \citealt{richeldi2022trial,chen2024efficacy}). 
In this paper, we will consider the specification by \cite{schmidli2014robust} with a slight difference that while the informative component in their case is a Meta-Analytic-Predictive (MAP) prior (obtained as the predictive distribution of a new study from a Bayesian meta-analysis of multiple external studies), in our case we consider the prior as coming from a single external study. This mixture prior, for a parameter of interest $\theta$, is given as follows:
\begin{equation}
	\pi_{\text{mixture}}(\theta)=w\pi_{\text{ext}}(\theta)+(1-w)\pi_{\text{robust}}(\theta),  
 \label{eq:1}
\end{equation}
\noindent
where $\pi_{\text{ext}}(\theta)$ is an informative prior originating, e.g. as a posterior distribution for $\theta$ from the analysis of a previous clinical trial with a noninformative prior, while $\pi_{\text{robust}}(\theta)$ is a weaker robust component whose role is to "dilute" the impact of the informative prior in the case that the informative prior is conflicting with the current data. Moreover, $w \in[0,1]$ is the prior weight of the external data. Such a formulation is akin to model averaging, as noted for example in  \cite{best2021assessing} and \cite{rover2019model}, where the two components represent the prior beliefs corresponding to whether external information is consistent with the current data or completely unrelated to the current data. The mixture weight then denotes the prior probability for each model. This prior is then combined with the current data, $y$, to obtain a posterior distribution. When the prior components are conjugate to the likelihood, the posterior is also a mixture given as follows:
\begin{equation}
	\pi_{\text{mixture}}(\theta|y)=\tilde{w}\pi_{\text{ext}}(\theta|y)+(1-\tilde{w})\pi_{\text{robust}}(\theta|y), 
 \label{mix_posterior}
\end{equation}
\noindent
where $\pi_{\text{ext}}(\theta|y)$ and $\pi_{\text{robust}}(\theta|y)$ are the respective posterior distributions under the individual prior components in (\ref{eq:1}).  The posterior weight, $\tilde{w}$, will determine the amount of "dilution" of the external data and updated as follows according to the data $y$ \citep{neuenschwander2023fixed}: 
\begin{equation}
\tilde{w}=\frac{wm_{\text{ext}}(y)}{wm_{\text{ext}}(y)+(1-w)m_{\text{robust}}(y)}, 
\label{post_weight}
\end{equation}
\noindent
where $m_{\text{ext}}(y)$ and $m_{\text{robust}}(y)$ are the marginal likelihoods of the data under the two components of the mixture prior (\ref{eq:1}), respectively, computed as follows: 
$$m_{\text{i}}(y)=\int_{}^{}\pi_{\text{i}}(\theta)f(y|\theta)d\theta, \quad i \in (\text{ext}, \text{robust}). $$
This mixture prior induces dynamic borrowing via the weight update allowing to borrow most when the prior and data are deemed similar and least otherwise.

\subsection{Robust component is normally distributed}
\label{robust component}
In this paper, we will focus on trials with a normal endpoint and known variance.
We consider a normal distribution for the robust component of the mixture prior (\ref{eq:1}) as follows:
$\pi_{\text{robust}}(\theta)= N  (\mu_{\text{robust}},\sigma^2_{\text{robust}})$, 
where $\mu_{\text{robust}}$ is the location and $\sigma^2_{\text{robust}}$ is the variance of the robust component. We also consider a normally distributed informative component, i.e. $\pi_{\text{ext}}(\theta)= N  (\mu_{\text{ext}},\sigma^2_{\text{ext}})$ with $\mu_{\text{ext}}=\bar{y}_{\text{ext}}$, the mean of the external data. The mixture prior (\ref{eq:1}) and posterior for $\theta$ are then both mixtures of two normally distributed components with computations available in closed-form. 

The mixture prior has gained popularity as an intuitive and easy to use Bayesian dynamic borrowing method. While the informative component parameters are dictated by the external data, specification of the robust component of this mixture prior remains challenging. 
The robust component requires specification of its location, dispersion and functional form as well as mixture weight. Nonetheless, all these choices act together in determining the posterior and therefore, their impact has to be evaluated jointly. 
In this paper, we evaluate whether such choices lead to unwanted frequentist operating characteristics (OC) or lead to the posterior distribution becoming multimodal. Already known is that this component should not be too vague as this results in Lindley's paradox \citep{o2004kendall}, i.e. even though a very dispersed robust component might not seem influential, it will result in a high posterior weight being assigned to the informative component. This is problematic in case there is bias between the current and external data \citep{rover2019model,mutsvari2016addressing}. To deal with this issue, weakly informative priors have been proposed. For example, for normal endpoints, the unit-information prior has been proposed \citep{kass1995reference, schmidli2014robust}. It is a prior whose variance has been scaled such that it contains information equivalent to that provided by one observation. More specifically, the variance of the unit-information prior is given by 
$
    \sigma^2_{\text{unit-information}}=\frac{1}{I(\theta)}
$, 
where $I(\theta)$ is the Fisher information based on a single observation. The impact of such a prior on the posterior distribution is more pronounced in case of small current sample size. 
This will be shown in more detail later. 

We will also investigate impact of location of the robust component and we consider three choices: 
\begin{itemize}
\item $\mu_{\text{robust}}=\bar{y}_{\text{ext}}$, i.e. the robust component is assumed to be located at the same location as external data (e.g. \cite{mutsvari2016addressing, callegaro2023historical, best2025beyond}). This applies in both one-arm and hybrid control trials. 
\item $\mu_{\text{robust}}=\theta_0$, i.e. the robust component is assumed to be located at the border of the null hypothesis to reflect scepticism and doubt about an effect \citep{spiegelhalter2004incorporating, edwards2024using, best2021assessing}. This applies only in the one-sample testing framework. Locating the robust component at the border of the null hypothesis is not an option for the hybrid control setting because the border between the null hypothesis and alternative hypothesis in this case is a line, i.e. all values $\theta_c$ and $\theta_t$ for which $\theta_t=\theta_c$.
\item $\mu_{\text{robust}}=\bar{y}$ (or $\bar{y}_c$ in the hybrid control setting), i.e. the robust component is assumed to be located at the observed mean of the current (control) data. While this choice is present in the literature \citep{berger1986robust, bennett2018improving}, its performance and applicability in practical situations has not been systematically investigated.   
\end{itemize}

\subsection{Heavy-tailed robust component}
\label{t-distribution}
We also consider a framework where the robust component of the mixture prior has heavy tails. This ensures that in case of large bias between the current and the external data, the latter is fully discounted \citep{o2004kendall}. We assume a location-scale t-distribution for this component:
$\pi_{\text{robust}}(\theta)= lst(\mu,\tau,\nu)$, 
where $\mu$ is the location, $\tau$ is the scale parameter while $\nu$ represents the degrees of freedom which regulates how heavy the tails are. Suggestions on using a t-distribution for robustness have been made e.g. in  \cite{schmidli2014robust} and \cite{o2004kendall}. We consider a t-distribution with 3 degrees of freedom as this is the heaviest t-distribution having finite mean and variance. We consider the scale $\tau=\sigma^2_{\text{unit-information}}$ such that the variance would match the variance of the normal setting considered herein as $\nu \to \infty$. Sensitivity to this choice can also be investigated. Note that the t-distribution is equivalent to a Normal mixture for which the variance is inverse-gamma distributed \citep{gelman2013bayesian}.

Adopting a t-distribution for the robust component requires numerical approaches to estimate the posterior. For ease of calculations, we approximate the t-distribution with a mixture of normals using Gamma quantiles. Under this framework, the robust component in the mixture prior (\ref{eq:1}) is then also a mixture of normals. Consequently, the mixture prior (\ref{eq:1}) is of the form: 
\begin{equation}
	\pi_{\text{mixture}}(\theta)=w\pi_{\text{ext}}(\theta)+(1-w)\sum_{k}^{}w_k\pi_{\text{k}}(\theta), 
\end{equation}
where $k$ is the number of components in the mixture of normal distributions approximating the t-distribution and $\sum_{k} w_k=1$. 
The resulting posterior is again a mixture with weights updated as in \cite{schmidli2014robust}:  
$$\tilde{w}=\frac{wm_{\text{ext}}(y)}{wm_{\text{ext}}(y)+(1-w)\sum_k w_km_k(y)} \quad \text{and} \quad
\tilde{w_k}=\frac{w_km_k(y)}{\sum_k w_km_k(y)}.$$ 
Of note, however, is that one may need to try different number of components to see which approximates the t-distribution best. More components may be needed depending on the heaviness of the tails of the t-distribution. As a check, we also used the exact t-distribution specification and computed the posterior distribution using MCMC to evaluate the Type I error rate on a coarse grid.
\section{Metrics for evaluation of borrowing approach}
\label{s:metrics}
\subsection{Operating characteristics (OCs): TIE rate, power and RMSE}
Even though many of the borrowing methods used in adaptive clinical trials are Bayesian, frequentist operating characteristics like TIE rate (eq.\ref{eq:TIE_onearm}, eq.\ref{eq:TIE_twoarm}), power, RMSE (eq.\ref{eq:MSE}) are typically used to evaluate the performance of such borrowing methods and are customarily sought by regulatory agencies. 
We follow suit evaluating TIE rate, power and Standardized RMSE, where the Standardized RMSE is obtained by dividing the RMSE of the posterior mean under the mixture prior by the RMSE of the maximum likelihood estimate. Though borrowing from external data may improve efficiency of the current trial when no or little bias is present, and while robust methods including the mixture allow to discount potentially conflicting external information, it is in general not possible to obtain a ‘perfect’ borrowing mechanism, i.e. one where gains in frequentist operating characteristics (OCs) are uniformly achievable \citep{kopp2020power}. It is, however, possible to gain some power at the cost of a TIE rate inflation. The amount of TIE rate inflation and whether it can be capped are typically of interest when evaluating dynamic borrowing approaches and are also investigated in this paper. To make a fair comparison of the power with and without borrowing, we utilize the recommendation proposed by \cite{kopp2024simulating} where the power without borrowing is based on a calibrated test using as significance level the TIE rate obtained with borrowing. 
Bayesian metrics have also been suggested in the evaluation of Bayesian designs \citep{berry2010bayesian, best2025beyond} and we, therefore, additionally evaluate average TIE rate and power as outlined in Section \ref{s:Bayes}. 
\subsection{Bimodality of posterior distribution }
As aforementioned, with a mixture prior the posterior distribution is also a mixture and if the locations of the components of the posterior distribution differ strongly in relation to dispersion, multimodality in the posterior distribution could emerge. With a multimodal posterior distribution, some quantities used to summarize the posterior distribution may be questionable. For example, the highest posterior density intervals (HPDIs) may be disjoint when the multimodality of the posterior distribution is strong. In this context the use of shortest credible intervals, which would coincide with the HPD intervals for unimodal distributions, might be preferable. In general, however, strong bimodality reflects unresolved prior-data conflict with respect to the informative component which is only partially addressed by the robust component. It is therefore important to assess and investigate it. When considering mixture priors and posteriors with two components, we will focus on bimodality as such a distribution can have at most two modes if the individual components are unimodal.  
Such bimodality, however, may present in different intensities depending on how separated the two modes are. We assess the separation between the two modes by the minimum of the ratio of each mode's density to the antimode's density, where the antimode represents the point of lowest density between the two modes \citep{o2004kendall}, hereafter referred to as O'Hagan's Bimodality Metric (OBM):
\begin{equation}
    OBM=min(f(\theta_{\text{mode 1}})/f(\theta_{\text{antimode}}),f(\theta_{\text{mode 2}})/f(\theta_{\text{antimode}})). 
    \label{bimodal-metric}
\end{equation}
Other metrics that could be used to assess the intensity of the bimodality in the posterior distribution include the assessment of whether the HPDIs consists of two disjoint intervals, as mentioned earlier, and Hartigan's dip statistic \citep{hartigan1985dip}.

We note that when a t-distribution approximated by a mixture of normal distributions is used for the robust component, the prior is a mixture of more than two normal components and hence it is possible for the resulting posterior to have more than two modes. However, further details on that is beyond the scope of this paper and for now we only investigate multimodality when the prior comprises a mixture of two normally distributed components. 
\subsection{Propagation of the prior weight}
Choice of the prior weight is subjective and has led to discussions on the best way to obtain this weight. Several approaches have been suggested \citep{zhang2023adaptively, egidi2022avoiding}. 
In the partial extrapolation of treatment efficacy from adults to adolescents, \cite{best2021assessing} propose a tipping point analysis where the prior weight is chosen as the smallest weight such that treatment benefit is shown in the adolescent subgroup and is agreed to represent a reasonable degree of similarity. 
In this paper, we additionally investigate how the posterior weight changes with respect to the prior weight for the different parameter choices of the robust component in the mixture prior. This provides an understanding into the importance of the prior weight selection. 
\subsection{Sweet spot}
Unlike in the one-arm case, in the hybrid-control case there exists a region, with respect to the external and current data bias, where locally power gain is possible whilst also having lower than frequentist TIE rate. We follow \cite{viele2014use} in terming such a region the "sweet spot". This region is characterized by no to low conflict between the current and external data and therefore there is no to small bias. However, this bias is compensated by reduction in variance. This leads to increase in the effective sample size of the control arm leading to lower TIE rate as well as higher power. We investigate the impact of the different parameter choices of the robust component on the sweet spot. In addition to the width of this spot, the maximum local power in this region is investigated.  
\section{Borrowing in a one-arm trial}
\label{s:one_arm}
\subsection{Problem set-up}
\label{one_arm_setup}
As noted earlier, we consider a normal endpoint. We assume the data $\bar{y}|\theta,\sigma \sim N(\theta,\sigma/\sqrt{n})$ and further assume $\sigma$ known. The hypothesis of interest is $H_0:\theta\leq \theta_0 $ versus $H_1:\theta>\theta_0$. For a prior for $\theta$, we consider the mixture prior (\ref{eq:1}) with normal components, i.e. 
$
    \pi_{\text{ext}}(\theta)= N (\bar{y}_{\text{ext}},\sigma^2_{\text{ext}})$ and $ \pi_{\text{robust}}(\theta)= N  (\mu_{\text{robust}},\sigma^2_{\text{robust}}). 
$ Additionally considered is the scenario where $\pi_{\text{robust}}(\theta)$ is t-distributed as introduced is Section \ref{t-distribution}. We evaluate TIE rate, Standardized RMSE and bimodality of the posterior distribution. In the one-arm trial setting considered, the test induced by borrowing is a uniformly most powerful test just at a modified TIE rate and corresponding power function. Hence there is a one-to-one relationship between power and TIE rate and this is the reason that reporting power is redundant in this case \citep{kopp2020power}. 

\subsection{Simulation set-up}
Without loss of generality we set $\sigma^2=1$. For this study, we take $n=20$ to reflect the small sample sizes where these methods are mostly used and consider an external study of size $15$ ($n_{\text{ext}}=15$) to avoid prior domination. Further, $\sigma^2_{\text{ext}}=\sigma^2/n_{\text{ext}}=1/n_{\text{ext}}$. We evaluate the operating characteristics for varying values of the observed external data mean, $\bar{y}_{\text{ext}}$, extending from no bias to extreme bias in both directions. For the one-arm normal test without borrowing, a power of $0.609$ is obtained for $n=20$ at $\theta_1=0.5$ and $\alpha=0.025$, where $\alpha$ is the significance level of the frequentist (no borrowing) test.
We investigate the robust component location choices mentioned in Section \ref{robust component} and for each location, we further investigate a range of variances such that the robust component spans from approximately flat to worth $2$ patients. Specifically, we investigate $\sigma^2_{\text{robust}}=\sigma^2/n_{\text{robust}}=1/n_{\text{robust}}$ to be $400, 25, 4, 1, 0.5$ where $1$ represents the unit-information in our case. Here $n_{\text{robust}}$ is an effective sample size representing the amount of information the robust component is worth and it is at most two. Additionally, for each location and variance choice, we investigate different values of the prior weight, $w$, assigned to the external data. We delay commentary on the choice $\mu_{\text{robust}}=\theta_0$ in the estimation framework to the discussion. For investigating TIE rate, the current data are generated with $\theta=\theta_0$. In the computation of TIE rate, RMSE and power with borrowing, one million Monte Carlo simulations were used. 
 
\subsection{Results for one-arm trial}
\subsubsection{Frequentist Operating Characteristics}
\begin{figure}[ht!]
  \includegraphics[width=\textwidth,height=\textheight,keepaspectratio]{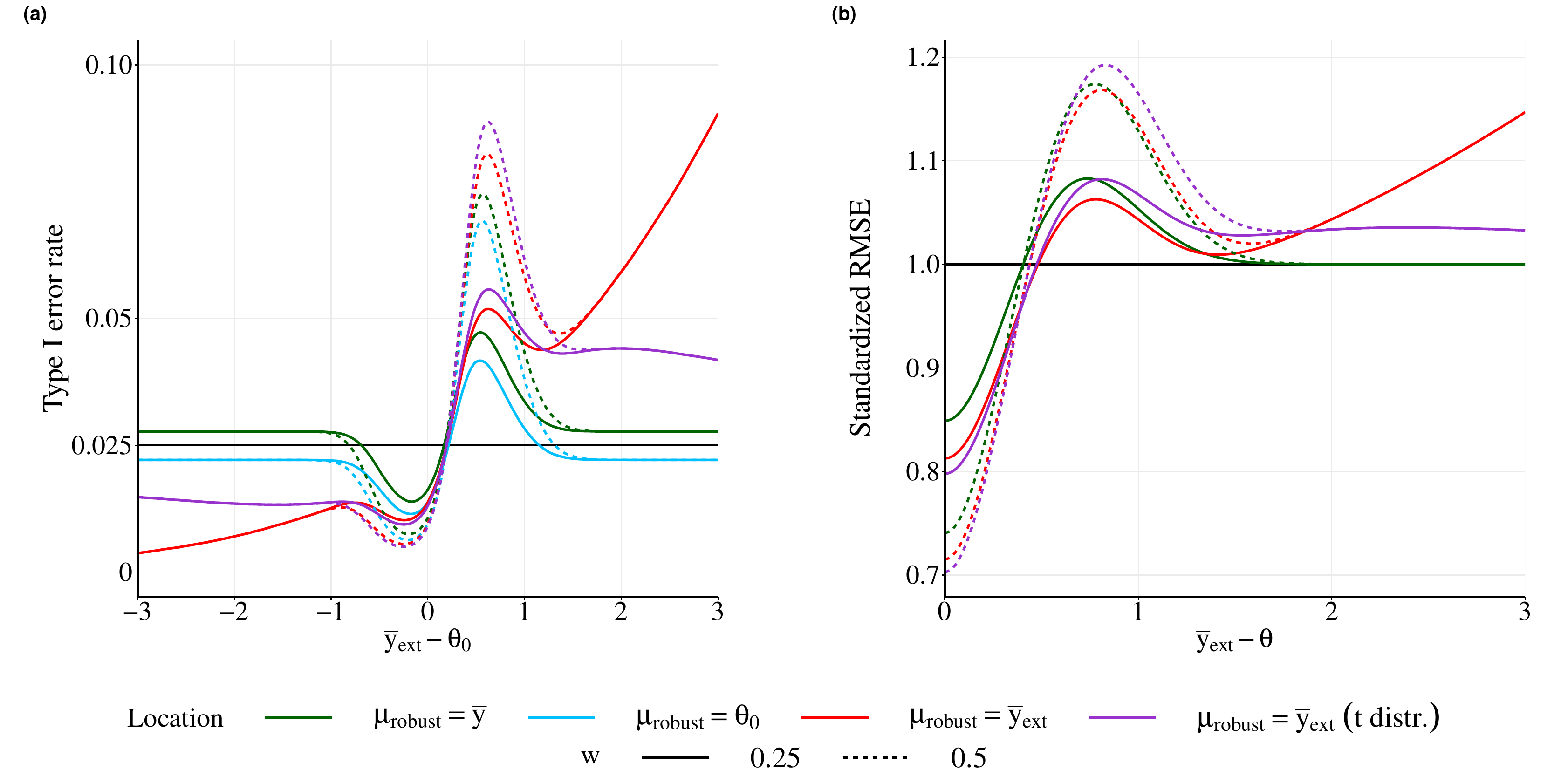}
	\caption{One-arm trial with the robust component of the mixture prior as a unit-information prior for $n_{\text{current}}=20$ and $n_{\text{ext}}=15$. Type I error rate (a) and Standardized RMSE (b) with the different robust component locations ($\mu_{\text{robust}}=\bar{y}_{\text{ext}}$, $\mu_{\text{robust}}=\theta_0$, $\mu_{\text{robust}}=\bar{y}$) represented by different colors. Different prior weights for the informative component are investigated $(0.25, 0.5)$ represented by the different linetypes. The purple lines represent the location $\mu_{\text{robust}}=\bar{y}_{\text{ext}}$ but with a t-distribution instead of a normal distribution. }
 \label{fig:TIE_MSE}
\end{figure}

Results of the TIE rate are shown in Figure \ref{fig:TIE_MSE} (a) for the different location scenarios and for selected prior weights, $w$, $0.25$ and $0.5$.  

In terms of dispersion, we focus on the unit-information prior as this is the current standard recommendation for the robust component. For the reasons explained in Section \ref{one_arm_setup}, results of power are not included in this figure but are nonetheless shown in the online Appendix Figure \ref{power_APPENDIX}. The x-axis represents the bias between the external data mean and the true mean of the current data with e.g. $0$, $2\sigma_{\text{ext}}$ and $8\sigma_{\text{ext}}$ representing no bias, moderate bias (approximately where OCs show a peak because external data is conflicting with the current but it has not been discarded yet) and large bias (where the external data has been discarded and the robust component takes over), respectively.
In all location scenarios, when there is mild bias in favor of the alternative hypothesis, TIE rate is lower than nominal level but with moderate bias TIE rate is inflated. This inflation also depends on the prior weight assigned to the external data. At extreme bias, TIE rate is the same regardless of the prior weight as the robust component tends to be assigned all posterior weight. The impact of the unit-information prior- which has information equivalent to that of one patient- will be more pronounced for small current sample size. Further results (Figures \ref{TIE_unit_ncurr}, \ref{TIE_rate_different_sample_sizes}, \ref{Power_different_sample_sizes}) show the impact of unit-information prior for the robust component for different current sample sizes and ratios of current to external sample sizes. Additional TIE rate results for different dispersions of robust component are also shown in Figure \ref{power_APPENDIX}. In that figure, importance of the location of the robust component is shown to decrease as this component's variance increases.

With $\mu_{\text{robust}}=\bar{y}_{\text{ext}}$, TIE rate inflation is higher compared to the other locations when there is moderate bias, and markedly so when there is extreme bias with the TIE rate eventually rising to one (not shown). However, with $\mu_{\text{robust}}=\theta_0$ or $\mu_{\text{robust}}=\bar{y}$, the upward trend in TIE rate is no longer present. As expected, choosing $\mu_{\text{robust}}=\theta_0$ leads to lower TIE rate, for $\bar{y}-\theta_0>0$,  than the other locations. 
On the other hand, choosing $\mu_{\text{robust}}=\bar{y}$ leads to artificial reduction of the posterior variance by one sample in the unit-information case and therefore a slight inflation in TIE rate compared to $\mu_{\text{robust}}=\theta_0$.  

Figure \ref{fig:TIE_MSE} (b) shows Standardized RMSE. The focus is again on the unit-information prior for the robust component, with varying prior weights, $w$. Here sensible locations for the robust component are $\mu_{\text{robust}}=\bar{y}_{\text{ext}}$ or $\mu_{\text{robust}}=\bar{y}$. With $\mu_{\text{robust}}=\bar{y}$, RMSE is lower than with $\mu_{\text{robust}}=\bar{y}_{\text{ext}}$ as we eliminate posterior bias. Additional results are shown in Figure \ref{MSE_APPENDIX} for different dispersions and prior weights of the robust component.

Figure \ref{fig:TIE_MSE} also shows TIE rate and Standardized RMSE when the robust component is t-distributed with location $\mu_{\text{robust}}=\bar{y}_{\text{ext}}$. The t-distribution is here approximated by a mixture of normal distributions with $100$ components. The upward trend in TIE rate observed when using a single normal distribution for this component is no longer present for the bias range considered. However, there is a slight inflation of TIE rate when there is moderate bias compared to only a single normal distribution for the robust component. Nonetheless, this slight inflation of TIE rate when using the heavier-tailed t-distribution might be acceptable given the significant reduction one achieves in the case that there is extreme bias. With extreme bias, where now all the weight is given to the robust component, using the exact t-distribution (based on MCMC) ensures that this prior is fully discarded and therefore the TIE rate/RMSE goes back to the frequentist level. This is a general characteristic of heavy-tailed priors, not just the t-distribution. We refer to \cite{ o2004kendall} and \cite{dawid1973posterior} for the theoretical proofs and further discussions. 

Selecting the number of components to approximate the t-distribution can pose a challenge. In Figure \ref{t_approx_APPENDIX}, we show the approximation of a t-distribution by a mixture of normal distributions with different number of components. The choice of the number of components is related to tail behaviour- the more components the better one captures the tails. In Figure \ref{number_of_components_onearm}, we show TIE rate results when the t-distribution is approximated by a mixture of normal distributions with different number of components as well as using the exact t-distribution (based on MCMC). The results show instability for low number of components while with many components the results are very close to MCMC. The results did not significantly change between $100$ and $200$ components, therefore we used $100$ components for all other computations. Convergence issues with the MCMC were sometimes observed at extreme bias (not shown) due to poor mixing of the chains. This may be attributed to the fact that at such extremes coupled with the heavy-tailedness of the t-distribution, the chains explore the space more slowly or get stuck. In Figure \ref{tdist_APPENDIX}, we show additional TIE rate results where the robust component is a t-distribution but with the scale $\tau$ calibrated such that the component has the same variance as the unit normal distribution. This
was done to make a better comparison of when the robust component is t-distributed versus normally distributed. This calibration results in a smaller scale leading to higher TIE rate inflation. 

\subsubsection{Bimodality of posterior distribution}
\begin{figure}[ht!]
\includegraphics[width=\textwidth,height=\textheight,keepaspectratio]{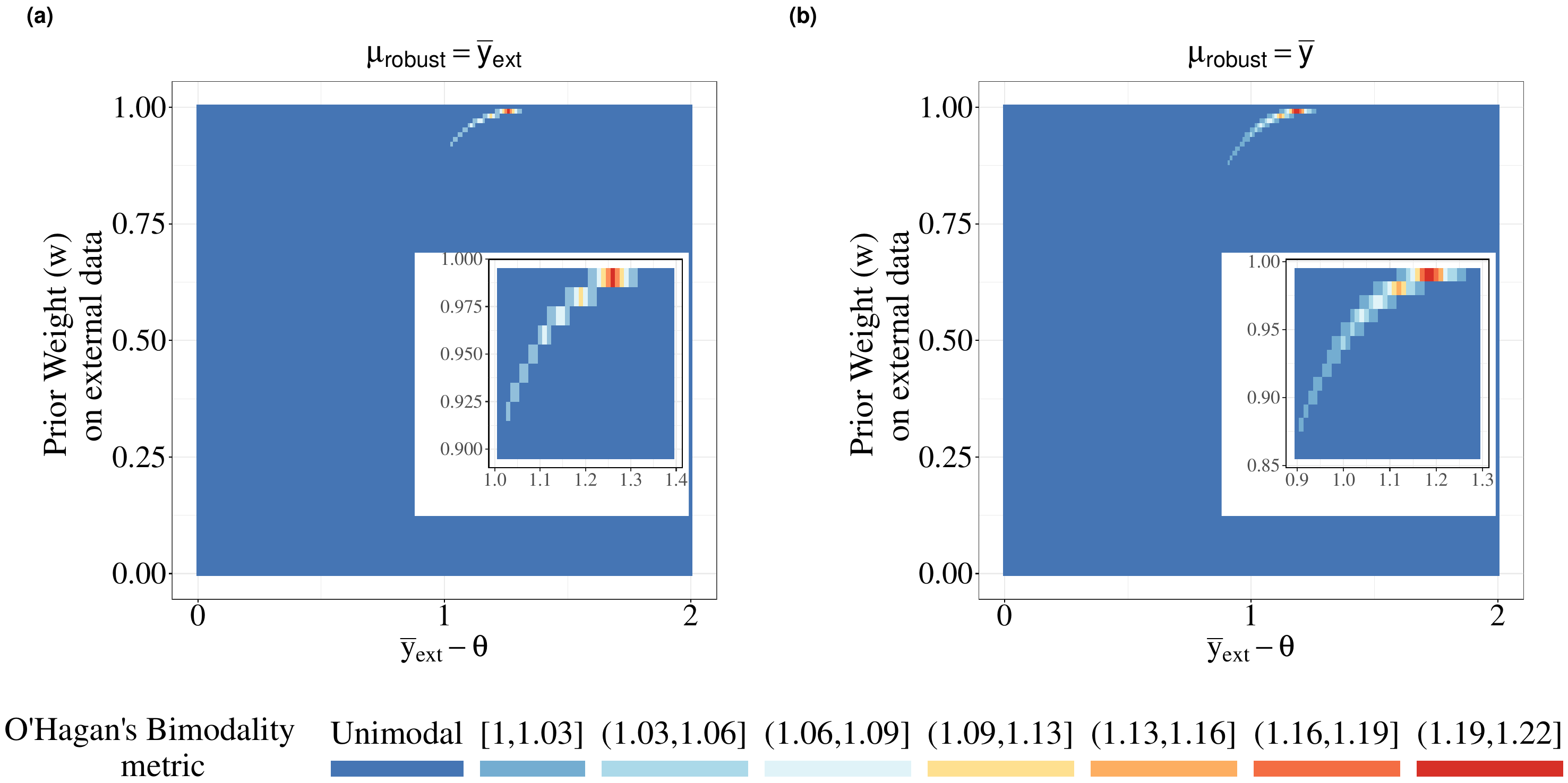}
\caption{One-arm trial with the robust component of the mixture prior as a unit-information prior. Bimodality intensity when $\mu_{\text{robust}}=\bar{y}_{\text{ext}}$ \text{(a)} and  $\mu_{\text{robust}}=\bar{y}$ \text{(b)} is shown. The y-axis represents a grid of prior weights, $w$, for the informative component. The x-axis represents the level of bias between the external data mean and the true mean of the current data. O'Hagan's Bimodality Metric (OBM) is used to measure the intensity of the bimodality. The inset zooms in on the region where bimodality is observed. Note $n_{\text{current}}=20 $ and $n_{\text{ext}}=15$.}
\label{bimodal_intensity}
\end{figure}

Figure \ref{bimodal_intensity} investigates the bimodality of the posterior distribution showing that under certain parameter choices, the posterior distribution can indeed be bimodal. OBM (\ref{bimodal-metric}) was used to evaluate how separated the two modes are. 
Results are shown for a grid of prior weights, $w$, and different external data means representing different bias ranges. The different robust component locations are also shown. The results are symmetrical around zero and therefore we only show results for positive bias. To obtain the posterior distributions, we set $\bar{y}=\theta$. More instances of bimodality in the posterior distribution are observed when $\mu_{\text{robust}}=\bar{y}$ compared to when $\mu_{\text{robust}}=\bar{y}_{\text{ext}}$. 
Additional results are shown in Figure \ref{bimodality_appendix} showing more instances of bimodality in the posterior distribution with increased robust component variance. For a unit-information robust component, bimodality is less likely with smaller prior weights and also with large current sample sizes relative to the external data sample size as shown in Figure \ref{bimodality_varying_n_appendix}. While there is no standard criterion for assessing the OBM, values greater than two can indicate high bimodality as the separation between the modes is then quite prominent. Bimodality in the posterior distribution can complicate decision making since there are now two regions of parameter values with high probability. For modes that are well separated, one may observe disjoint Highest Posterior Density intervals for commonly used intervals like $90\%$ or $95\%$.
\subsubsection{Posterior weights}
\begin{figure}[ht!]
\includegraphics[width=0.95\textwidth,height=0.9\textheight,keepaspectratio]{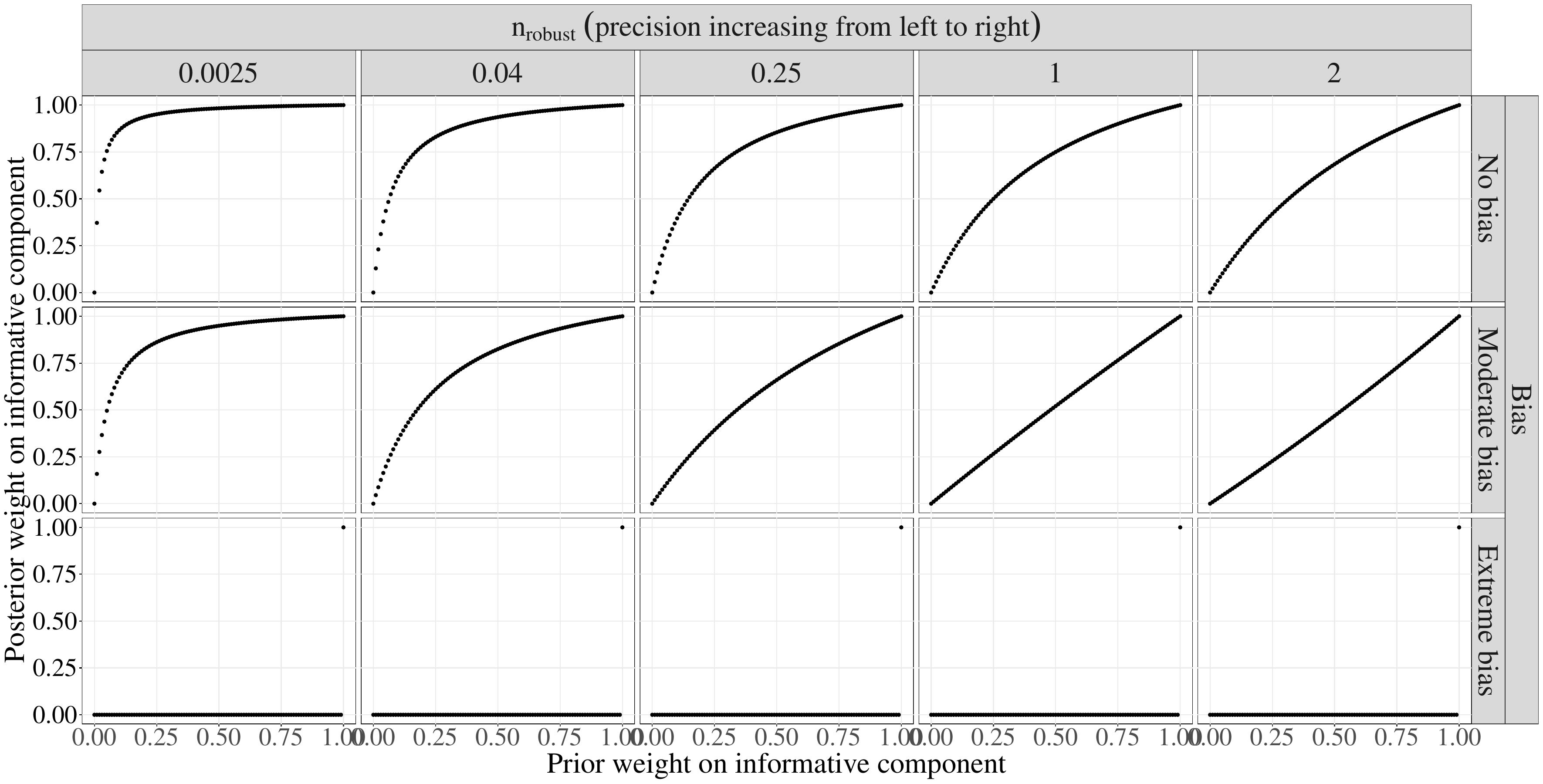}
	\caption{Posterior versus prior weight of informative component. The columns represent precision of the robust component, while the rows represent different levels of bias: $0$ (No bias), $2\sigma_{\text{ext}}$ (moderate bias) and $8\sigma_{\text{ext}}$ (extreme bias). The column $n_{\text{robust}}=1$ represents the unit-information. Here $\mu_{\text{robust}}=\bar{y}_{\text{ext}}$, $n_{\text{current}}=20$ and $n_{\text{ext}}=15$.} 
 \label{wtilda}
\end{figure}

\begin{figure}[ht!]
\centering
\includegraphics[width=\textwidth,height=\textheight,keepaspectratio]{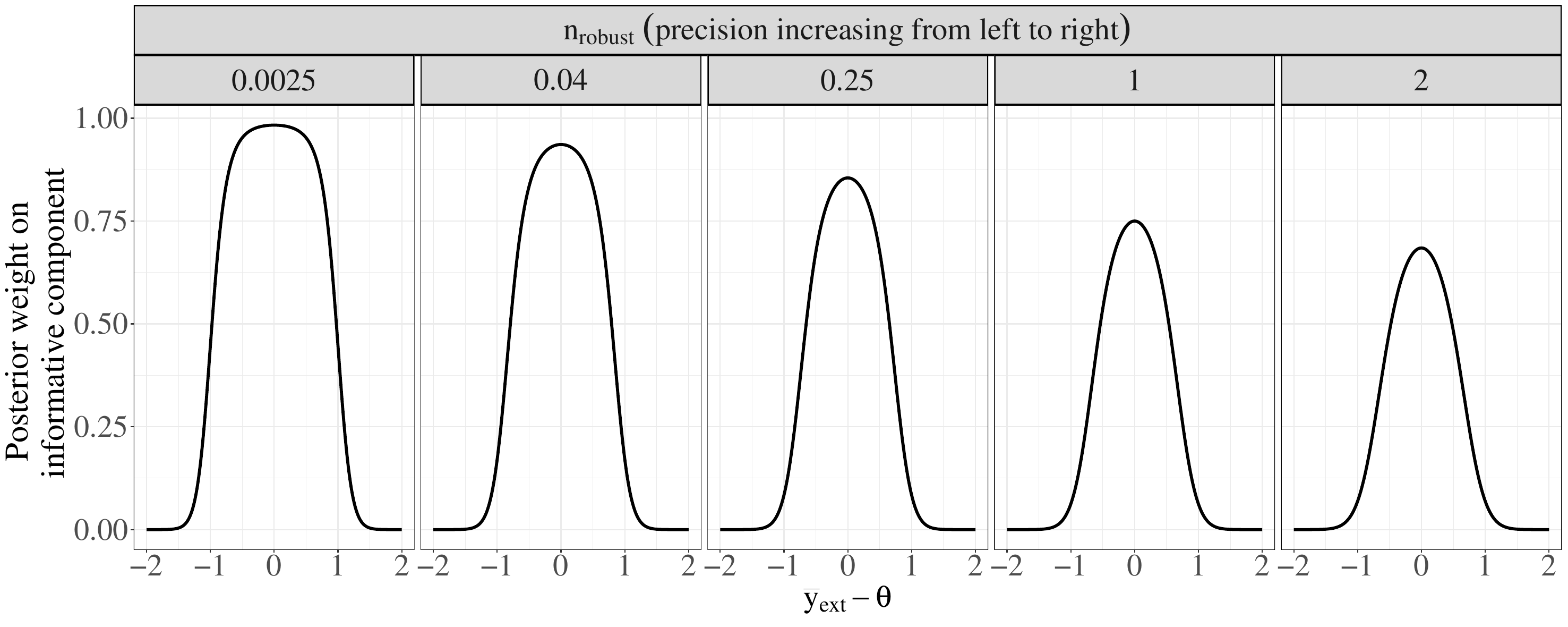}
	\caption{Posterior weight of informative component versus bias when prior weight $w=0.5$. The columns represent precision of the robust component, with $n_{\text{robust}}=1$ representing the unit-information. Here $\mu_{\text{robust}}=\bar{y}_{\text{ext}}$, $n_{\text{current}}=20$ and $n_{\text{ext}}=15$. } 
 \label{wtilda_w0.5}
\end{figure}

Figure \ref{wtilda} shows the resulting posterior weights for different prior weights and variances for the robust component. These are shown for different bias scenarios; namely, no bias ($0$), moderate bias ($2\sigma_{\text{ext}}$) or extreme bias ($8\sigma_{\text{ext}}$). Having a large variance for the robust component when there is no bias results in almost all the weight being allocated to the informative component, except for prior weight very close to one. Whilst this is the wanted situation, this benefit breaks down when there is moderate bias. In that case, a high posterior weight is still assigned to the informative component. This stems from the known Lindley's paradox. On the other hand, the unit-information robust component ensures that this prior adapts faster to bias. With extreme bias, a prior weight greater than zero is updated to one regardless of the robust component's variance. 
To emphasize the impact the robust component has in the updating of the weights in the mixture, we consider a prior weight of $0.5$ and display in Figure \ref{wtilda_w0.5} how the posterior weight changes depending on the level of bias between the current and external data and informativeness of the robust component. 
\section{Hybrid control trial}
\label{s:two_arm}
\subsection{Problem set-up}
\label{two-arm: set-up}
We consider a two-arm trial comparing a treatment group to control group with external information available to be borrowed for the control group. Again, we consider normally distributed endpoints with $\bar{y}_t|\theta_t,\sigma_t\sim N(\theta_t,\sigma_t/\sqrt{n_t})$ being the current treatment outcome and $\bar{y}_c|\theta_c,\sigma_c\sim N(\theta_c,\sigma_c/\sqrt{n_c})$ being the current control outcome. We further let $\sigma_t=\sigma_c=\sigma$. We test the following one-sided hypothesis: 
$
     H_0: \theta_t \leq \theta_c $ versus $H_1: \theta_t > \theta_c
$.
We consider the mixture prior (\ref{eq:1}) for $\theta_c$ with normally distributed components as in Section \ref{one_arm_setup} while a flat prior is used for $\theta_t$. As in the one-arm setting, an additional scenario is investigated where $\pi_{\text{robust}}(\theta)$ is t-distributed as introduced is Section \ref{t-distribution}.

\subsection{Simulation set-up}
We again consider $\sigma^2=1$ and $n_t=n_c=20$ for the current treatment and control sample sizes, with $n_{\text{ext}}=15$ as the external control sample size. We consider similar set ups for the robust component parameter choices as in the one-arm trial. TIE rate is computed under the scenario $\theta_t=\theta_c$. 
For the one-sided two-sample test without borrowing, a power of $0.75$ is obtained at $\theta_t-\theta_c=0.83$ and $\alpha=0.025$, where $\alpha$ is the significance level of the frequentist (no borrowing) test.  
\subsection{Results for Hybrid control trial}
\begin{figure}[ht!]
   \centering
  \includegraphics[width=\textwidth,height=\textheight,keepaspectratio]{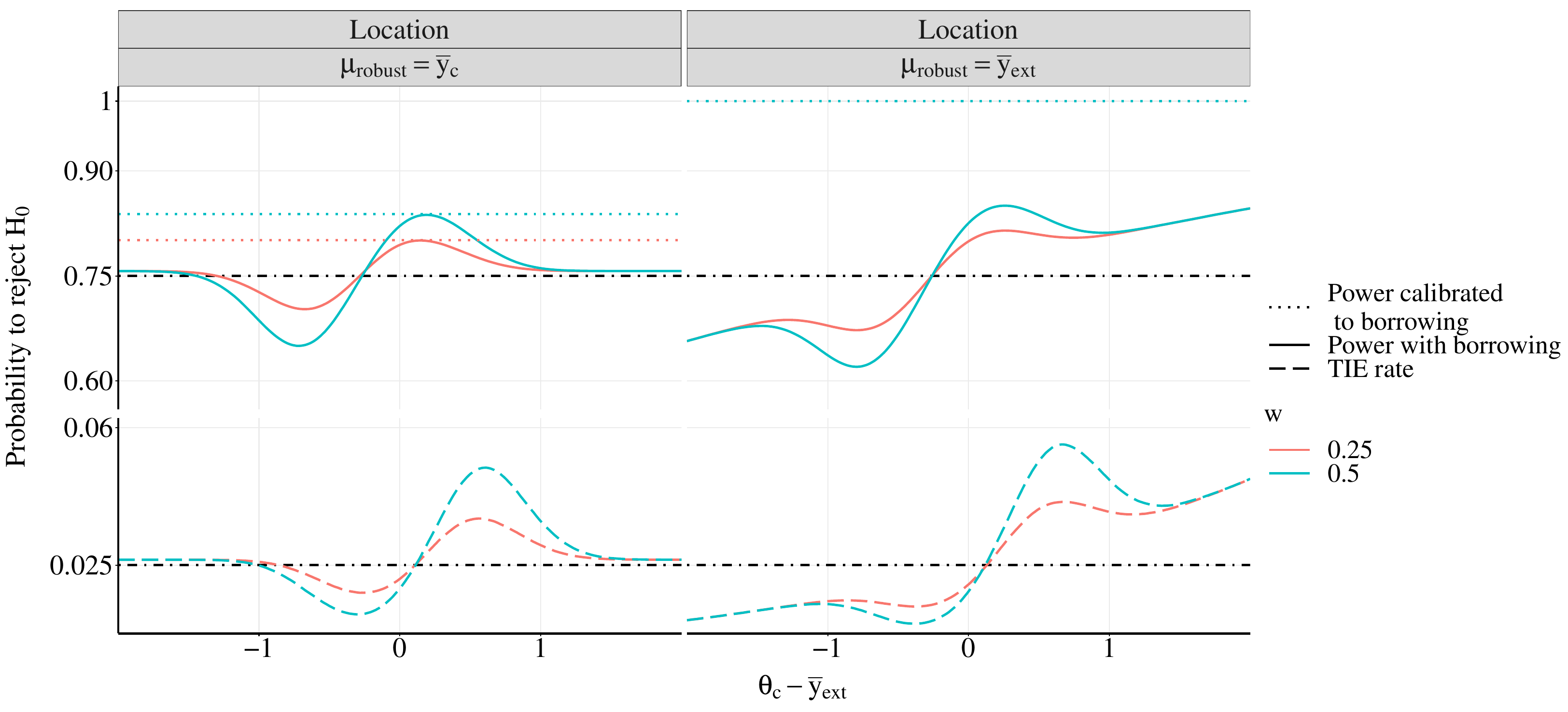}
	\caption{Hybrid control trial with the robust component of the mixture prior as a unit-information prior. Power with borrowing (solid lines), TIE rate (dashed lines) as well as the power calibrated to borrowing (dotted lines) are shown for the two robust component locations and different prior weights. The dotdash lines represent the frequentist power and TIE rate, at $0.75$ and $0.025$, respectively. Here, $n_t=n_c=20$ and $n_{\text{ext}}=15$. The frequentist power of $0.75$ is obtained at $\theta_t-\theta_c=0.83$ and $\alpha=0.025$. }
 \label{power_tie:2arm}
\end{figure}

Results of TIE rate and power are shown in Figure \ref{power_tie:2arm} for a unit-information robust component. Unlike in the one-arm case where TIE rate and power are directly related, in the hybrid-control setting the relationship also involves the unknown current control mean $\theta_c$. Depending on the informativeness and location of the robust component, as in the one-arm setting, it is possible for the TIE rate to go to one. Therefore, when computing the power of the test calibrated to borrowing, it is essential to evaluate TIE rate for all $\theta_c-\bar{y}_{\text{ext}}$ and to calibrate to the maximum TIE rate \citep{kopp2024simulating}. As such, when $\mu_{\text{robust}}=\bar{y}_{\text{ext}}$, the power of the test calibrated to borrowing is calibrated not to the maximum TIE rate in the shown range but to the maximum TIE rate in the full range, i.e. one. In Figure \ref{power_tie:2arm}, the power of the test calibrated to borrowing is indicated by the dotted lines. The difference in the power of the test with borrowing and calibrated to borrowing is shown in Figure \ref{power_diff:2arm}. This difference is always negative, sometimes small, but never zero. 

\begin{figure}[ht!]
\includegraphics[width=0.85\textwidth,height=0.8\textheight,keepaspectratio]{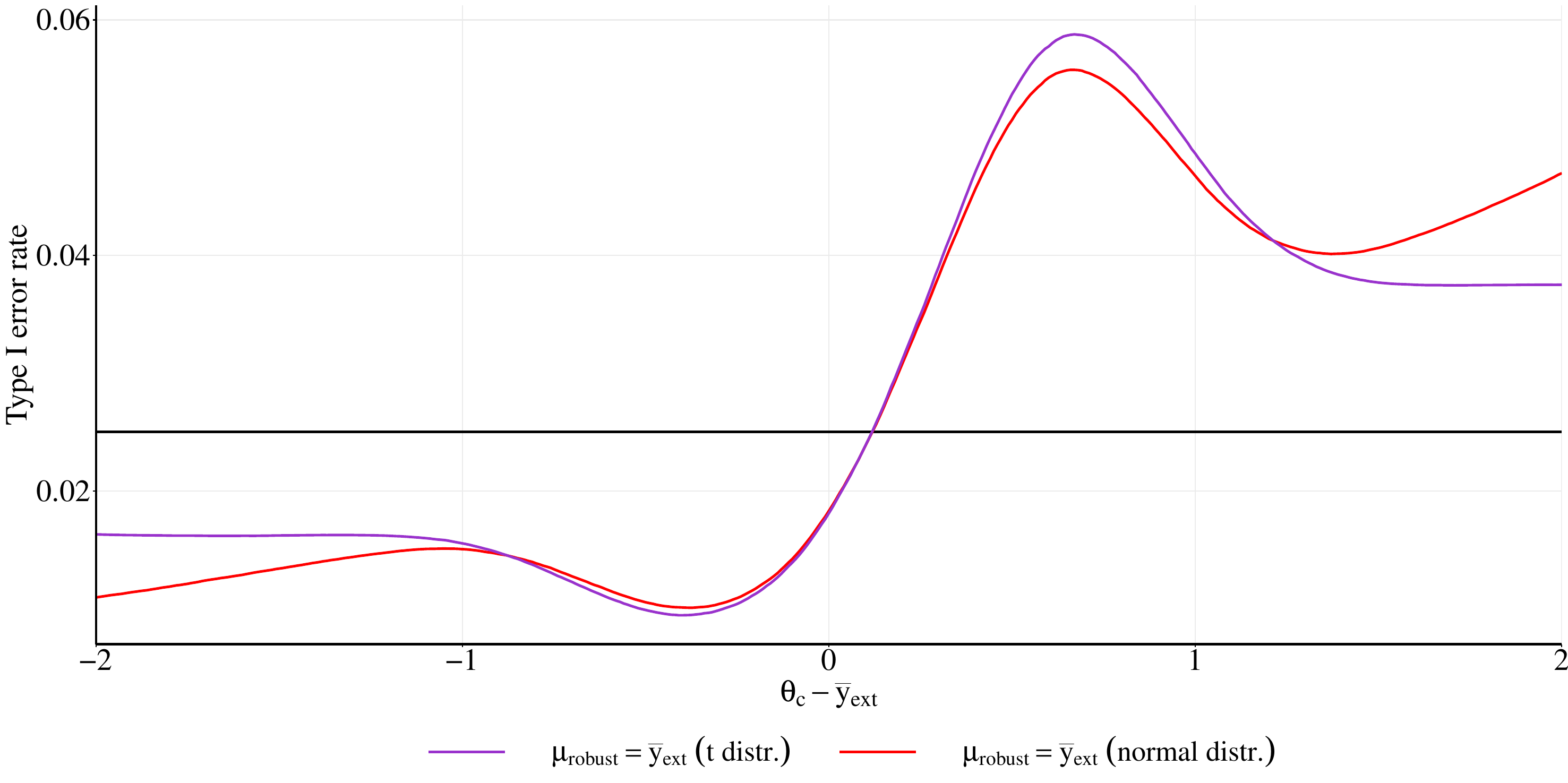}
	\caption{TIE rate for Hybrid control trial where the robust component of the mixture prior is a normal distribution with unit information or t-distributed with scale $1$ and three degrees of freedom. Here, $\mu_{\text{robust}}=\bar{y}_{\text{ext}}$, $w=0.5$, $n_t=n_c=20$ and $n_{\text{ext}}=15$. The t-distribution is approximated by a mixture of normals with $100$ components.  }
 \label{tie_t:2arm}
\end{figure}

As in the one-arm trial, we also investigate TIE rate when the robust component is t-distributed with location $\mu_{\text{robust}}=\bar{y}_{\text{ext}}$.  Again, the t-distribution is approximated by a mixture of normal distributions with $100$ components. This is shown in Figure \ref{tie_t:2arm} for a prior weight of $0.5$ alongside the TIE rate assuming a normal distribution for the robust component. Similar conclusion as in the one-arm trial is drawn where for large bias, using a t-distributed robust component avoids the upward trend in TIE rate seen with the normal distribution.  

As noted in \cite{viele2014use}, in the two-arm case, a "sweet spot" does exist where gains in power and reductions in TIE rate are possible. This sweet spot covers values for bias $\theta_c-\bar{y}_{\text{ext}}$ for which TIE rate is at most $\alpha$, and power is larger or equal to the power of the test without borrowing. Figure \ref{sweetspot_illustration} shows an example of a sweet spot for $w=0.5$ with a unit-information robust component with $\mu_{\text{robust}}=\bar{y}_{\text{c}}$.   
\begin{figure}[ht!]
\centering\includegraphics[width=\textwidth,height=\textheight,keepaspectratio]{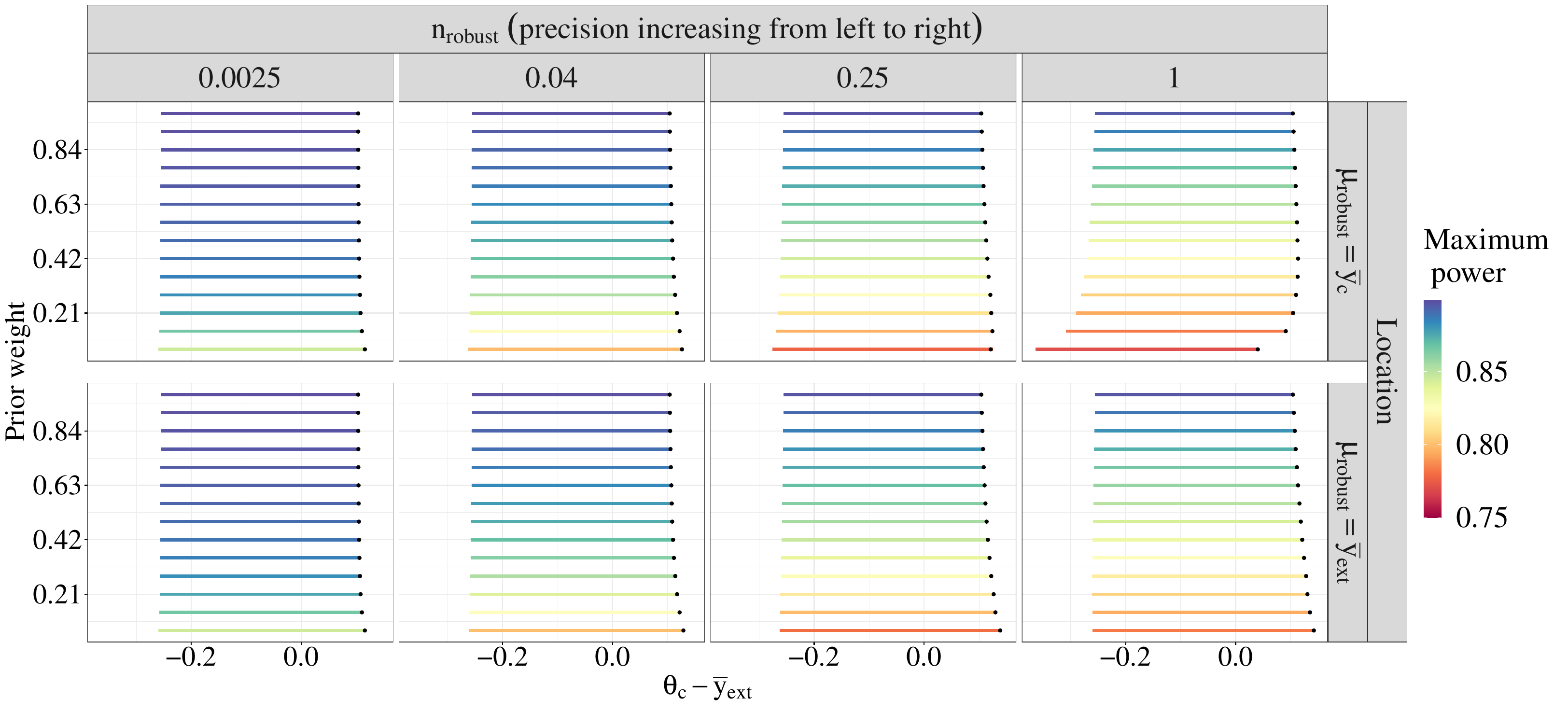}
	\caption{In the Hybrid control trial, the size of the sweet spot is shown for the different locations and dispersions for the robust component and different prior weights assigned to the informative component. The rightmost column represents the unit-information. The color indicates the maximum power in the sweet spot while the dot shows where in this spot the maximum power is observed. For contrast, the one-sided two-sample test without borrowing has a power of $0.75$ at $\theta_t-\theta_c=0.83$ for $n_t=n_c=20$ and $\sigma^2=1$ at $\alpha=0.025$.}
 \label{sweetspot:2arm}
\end{figure}
The size of the sweet spot is impacted by not only the prior weight assigned to the informative component but also by the informativeness of the robust component and size of external data. 
Figure \ref{sweetspot:2arm} shows the range of this sweet spot for different locations and dispersion of the robust component and different prior weights. Additionally, the maximum power in the sweet spot is shown as well as where in this spot the maximum power occurs. 
Differences in the size of the sweet spot between the two robust component locations are more pronounced for small prior weights with the spot being shifted more to the left when $\mu_{\text{robust}}=\bar{y}_{\text{c}}$ compared to when $\mu_{\text{robust}}=\bar{y}_{\text{ext}}$. 
The smaller the prior weight assigned to the informative component, the wider the sweet spot. However, less difference is observed between the prior weights as the robust component becomes more vague. 
\begin{table}[ht!]
\centering
\caption{Maximum TIE rate (\%) and Maximum power gain (\%) in the hybrid-control trial where the maximum size of bias between the external and current data is restricted by $\Delta$, i.e. $\lvert \theta_c-\bar{y}_{\text{ext}}\rvert\leq \Delta$. Here $n_t=n_c=20, n_{\text{ext}}=15, n_{\text{robust}}=1$ and $w=0.5 $. For comparison, the test without borrowing has TIE 2.5\% and power 75\%.}
\label{Table:Delta}
\begin{tabular}{cccll}
\hline
\multicolumn{1}{l}{} & \multicolumn{2}{c}{$\mu_{\text{robust}}=\bar{y}_{\text{ext}}$}      & \multicolumn{2}{c}{$\mu_{\text{robust}}=\bar{y}_{\text{c}}$}                                            \\
\hline
$\Delta $               & Max TIE rate (\%) & Max power gain (\%) & \multicolumn{1}{c}{Max TIE rate (\%)} & \multicolumn{1}{c}{Max power gain (\%)} \\
\hline
0.1                  & 2.38         & 9.79           & 2.43                             & 8.79                               \\
0.2                  & 3.08         & 7.15           & 3.08                             & 5.95                               \\
0.4                  & 4.57         & 2.26           & 4.39                             & 1.43                               \\
0.5                  & 5.15         & 0.82           & 4.82                             & 0.29                              \\
\hline
\end{tabular}
\end{table}
Moreover, the maximum power seems to be inversely related to the width of the sweet spot with higher power observed if the sweet spot is narrow compared to when it’s wider. For most of the cases shown, the maximum power was observed at the edge of the sweet spot though for some the maximum power is observed slightly before the edge. Larger prior weights are also associated with higher maximum power in this spot. Additional results are shown in Figure \ref{a:sweetspot} illustrating what happens to the size of the sweet spot when the current trial sample size changes. 

In Figure \ref{power_tie:2arm}, the power of the test calibrated to borrowing was calibrated to the maximum TIE rate over the full range. Nonetheless, if there is certainty that the maximum size of bias is within some $\Delta$ around $0$, then one may calibrate to the maximum TIE rate in this range \citep{stall2020}. With such a restriction, a power gain is possible. In Table \ref{Table:Delta}, we show an example for a prior weight of $0.5$ with a unit-information robust component. The smaller the $\Delta$, i.e. the more confident we are that the bias is small, the higher the power gain.

We additionally consider a setting where the unit-information component used in the mixture prior for the control arm, is used as prior for the treatment arm with location $\mu_{\text{robust}}=\bar{y}_{\text{ext}}$ as done in RBesT \citep{weber2019applying}. Figure \ref{unit:treatarm} shows the resulting TIE rate. Compared to the right-side plot of Figure \ref{power_tie:2arm}, the TIE rate inflation is controlled when there is extreme bias. This controlled behavior is, however, only observed when the two arms have the same sample size. With unbalanced sample sizes, the uncontrolled inflation resumes as seen in Figure \ref{unit_treatarm_unbal}.

\begin{figure}[ht!]
   \centering
  \includegraphics[width=\textwidth,height=\textheight,keepaspectratio]{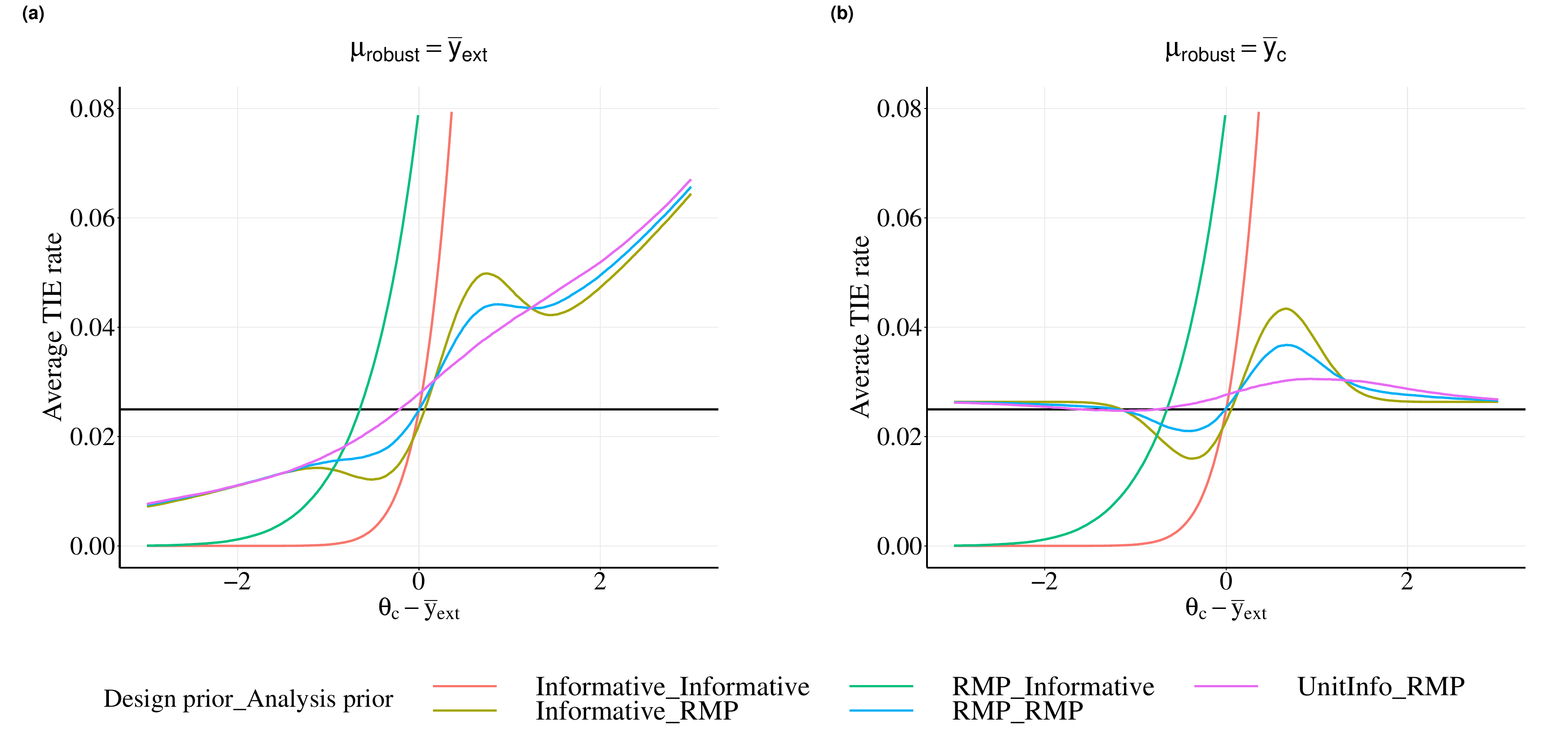}
	\caption{Average TIE rate for the hybrid control trial setting is shown for different design and analysis priors. The analysis prior is additionally shifted to reflect a range of prior-data conflict. (a) and (b) represent the two different robust component locations for the robust mixture analysis prior, i.e. $\mu_{\text{robust}}=\bar{y}_{\text{ext}}$ and  $\mu_{\text{robust}}=\bar{y}_{\text{c}}$, respectively. The legend shows the respective priors separated by underscore (i.e. design prior\_analysis prior). Where the robust mixture prior (RMP) is shown, a weight of $0.5$ was used. Here $n_t=n_c=20$ and $n_{\text{ext}}=15$.} 
 \label{average_tie:2arm}
\end{figure}

For the evaluation of average TIE rate and power, as outlined in Section \ref{s:Bayes},  design and analysis priors are needed. In analogy to \cite{best2025beyond}, we consider three choices for the design prior:
\begin{itemize}[itemsep=0pt, topsep=0pt]
    \item Informative prior based on the external data (Informative), corresponding to $w=1$ in the robust mixture prior in \eqref{eq:1}.
    \item Robust mixture prior (RMP), with weight $0.5$ in \eqref{eq:1}. 
    \item Robust component of the mixture prior, i.e. the unit-information robustifying the informative component in the mixture prior (UnitInfo), corresponding to $w=0$ in \eqref{eq:1}.
\end{itemize}
As analysis prior we consider again the above three choices but additionally shift the analysis prior to reflect a range of prior-data conflict. For the robust mixture prior as analysis prior, we also evaluate the two robust component locations. The average TIE rate results are shown in Figure \ref{average_tie:2arm} while the average power results are shown in Figure \ref{average_power:2arm}. As noted in \cite{best2025beyond} when the design prior and analysis prior match, average TIE rate is controlled at the pre-specified level $\alpha=0.025$. In Figure \ref{average_tie:2arm}, that means average TIE rate at $0.025$ for no bias. With the robust mixture as analysis prior, similar conclusions are observed as in the classical TIE rate setting considered above: adopting $\mu_{\text{robust}}=\bar{y}_{\text{c}}$ avoids the upward trend in average TIE rate. 

\section{Discussion} 
\label{s:discuss}
\subsection{Summary}
When using the robust mixture prior for borrowing from external data, parameter choices have to be made. This is non-trivial for the parameters of the robust component of the mixture prior as well as the mixture weights. All three parameter choices are influential, acting together and therefore their impact needs to be assessed jointly.

Regarding the location of the robust component, we have investigated the three feasible options: at the border of the null hypothesis in case of testing, same as external data mean and at the current data observed mean. In the hybrid-control trial, only the last two are plausible as locations for the robust component since, as mentioned before, the border between the null and alternative hypothesis in this case is a line (i.e. $\theta_t=\theta_c$). 
We investigated a range of dispersions ranging from a very vague robust component to the unit-information, which is the current recommendation.  Additionally, different values for the prior weight were considered. For both, one-arm and hybrid-control trials, TIE rate and power were evaluated. Additionally, for the hybrid-control setting, the "sweet spot" was investigated, where control of TIE rate and gain in power is observed locally. The broadness of this region can be used to gauge the choice of parameters of the robust component. The other metrics assessed were Standardized RMSE and multimodality of the posterior distribution.

We found that for some parameter choices, losses may still be unmitigated despite the use of dynamic borrowing for both testing and estimation, i.e. TIE rate may rise to one while RMSE may increase unconstrained. As shown before, no power gains are possible if one wishes to control the TIE rate to the nominal level if a UMP test exists \citep{kopp2020power}. In the hybrid-control trial, a sweet spot can be found but this only relates to a specific range of current and external data bias. The nice feature of mixing two prior distributions gives flexibility but comes at the cost of potential bimodality in the posterior distribution, which might complicate decision making.

\subsection{Recommendations}
\label{recommendations}
In Table \ref{Table:recommendations}, we have provided a summary of some of the recommendations discussed herein.
\subparagraph{Location}
Recommendation for the choice of location for the robust component will depend on whether one is in the testing or estimation framework. In the one-arm testing framework, locating a normally distributed robust component at the external mean is the only option that did not ensure a cap on TIE rate inflation while all the other options that we considered did. For the location in hybrid-control testing, the border between the null and alternative hypotheses is not an option as it is a line (i.e. $\theta_t=\theta_c$). Here, locating the robust component at the external control but with a heavy-tailed distribution will ensure a cap on the TIE rate inflation. Alternatively, locating the robust component at the current control observed mean will also ensure such a cap in TIE rate. Due to the artificial variance deflation that occurs when locating the robust component at the current control observed mean, TIE rate does not go back to exactly the frequentist one, though differences are typically minor, and they are bounded. Locating the robust component at the current control observed mean makes the prior data-dependent - contrary to a fully Bayesian perspective that avoids double use of the data - and can be thought of as an empirical-Bayes approach. Such methods where the prior choice is guided by the observed data have a long history, some early references include \cite{morris1983parametric}, and \cite{berger1986robust} for the epsilon-contaminated/mixture class of distributions, and references therein. In the context of dynamic borrowing for clinical trials, Empirical Bayes versions have been proposed for both the power prior \citep{gravestock2017adaptive} and the commensurate prior \citep{hobbs2012commensurate} approach. 
In general, the main drawback of empirical-Bayes approaches is that they do not guarantee Bayesian optimality, but they often lead to estimators with good frequentist behavior \citep{robert2007bayesian} and we have not encountered any problematic behaviour in our simulations.

If estimation is required in addition to testing, locating the robust component at the border of the null hypothesis would introduce shrinkage in the estimate towards a null effect. An unbiased estimate for large bias is, however, obtained by locating the robust component at the current data observed mean. Nonetheless, the above concerns as in the testing framework apply.
\subparagraph{Dispersion}
Despite the unit-information being the current recommendation for the dispersion of the robust component \citep{schmidli2014robust}, if the current sample size is small, the impact of the unit-information prior on the posterior will be more pronounced. As we have seen in our results, this can lead to undesirable operating characteristics for some choices of the other parameters of the robust component. This is critical as such borrowing methods are mainly used with small populations. In case of large sample size, the impact would only be relevant for very large bias. We recommend to run simulations to check the actual impact that the unit-information has, especially when the sample size is small. Nonetheless, locating the unit-information robust component at the current data mean or adopting a heavy-tailed distribution does remedy the undesirable OCs. With either of the two choices, both TIE rate and RMSE are capped. We have shown that the importance of the robust component's location decreases with increasing variance. However, since this component should not be very vague as this leads to Lindley's paradox, the choice of location remains important.
\subparagraph{Functional form}
The heavy tails enhance the robustness of the mixture prior. Admittedly, however, the need to choose the scale parameter and degrees of freedom adds a layer of complexity. Adopting a t-distribution would require running MCMC but this can be somewhat simplified by approximating by a mixture of normal distributions, though one still needs to decide on the number of components of said mixture. One could consider concurrently running MCMC for a few points, covering a range of bias, which would give a rough idea on the number of components needed for the approximation without being too computationally intensive. Nonetheless, for any finite number of Gaussian components, the tails of the approximation will always be Gaussian and therefore such an approximation will always be less robust than using a real t-distribution. 
\subparagraph{Weight}
Choices for the weight should first of all be based on considerations of plausibility of current and external data congruence. One would still inspect the OCs to evaluate what might happen if the pre-specified choice was flawed or when one does not have an initial clue. As expected, we have seen in our results that lower weights lead to smaller gains but also less dramatic losses. A tipping point analysis may also assist in selecting the weight \citep{best2021assessing}. While assigning a prior distribution to the weight has been suggested instead of considering the weights fixed, \cite{neuenschwander2023fixed} shows that this does not impact inference for the parameter of interest as identical results would be obtained if the fixed weights are equal to their prior means.
\subparagraph{Evaluation of OCs/simulation}
To allow for a fair comparison of power with and without borrowing, the power without borrowing should be based on a calibrated test with significance level as the TIE rate obtained with borrowing as proposed by \cite{kopp2024simulating}. Whenever the TIE rate is rising to one, the power calibrated to borrowing is at the maximum possible value of one. In the hybrid control case, it is important to stress that TIE rate should be computed over all possible values for current and external data bias since $\theta_c$ is unknown. Calibrating to the maximum TIE rate in a restricted range instead of the full range may lead to an overoptimistic conclusion of power gain.
\subparagraph{Prior domination}
One might encounter a case where external data dominates or overwhelms the current data, for example if a large trial on adults is available for use to inform a much smaller pediatric trial. To mitigate overwhelming of the current data by the external data, setting the variance of the informative component of the mixture prior to the current data variance ensures that  borrowed information is not more than the current one \citep{chen2006relationship}. An alternative would be to use a meta-analytic approach, with an appropriately chosen prior for the between-study heterogeneity parameter, to formulate the informative prior from the external data as a heavy-tailed predictive distribution. Nonetheless, the resulting MAP prior would still need to be robustified for prompt adaptation to bias between the external and current data \citep{schmidli2014robust}.   
\section{Conclusion}
\label{s:conclusion}

Borrowing external data for use in a current study has emerged as an attractive research area with potential to make current studies more efficient especially where recruitment of patients is difficult. We have investigated one-arm and hybrid control trials with normal endpoints and fixed external data. Other scenarios can occur. Sometimes external data is available on the treatment effect in a two-arm trial, in which case the assessment boils down to a one-arm setting and therefore the evaluations herein regarding borrowing in a one-arm trial apply directly. Although here we considered borrowing from external trial data, such information may also be from other external sources e.g. real world or expert opinion. Additionally, although our findings would apply under normal approximation of e.g. logit-transformation for binary endpoints or log-hazard ratio for time-to-event endpoints, further exploration of alternative models may be needed and will be investigated in the future. Note that when the normal approximation is applied to log-odds or log-hazards, proposals for the robust component specification may follow from proposals for weakly informative prior distributions in these contexts. For example, for log-odds, a Normal($0, \pi^2/3$) prior \citep{rover2021weakly}, mimicking a uniform prior in probabilities, may be an option for the robust component.

We investigated here the situation of external data that is fixed and known at the design stage of the trial. Situations exist in which external data should rather be considered as random, i.e. not yet available or selected from a larger dataset. Existing research \citep{kopp2024simulating} suggests that the operating characteristics will most likely be similar but potentially less favorable. 

Bias between the current and external data is an issue and may arise because of several reasons. Pocock's criteria \citep{pocock1976combination} or experts knowledge can offer insights on predictable sources for such bias and can guide the initial selection e.g. of the borrowing weights. The robust mixture prior and other dynamic borrowing approaches provide additional robustness against current and external data bias from unknown sources. As we have investigated in this work, additional careful thought needs, in general, to be given to the choice of the tuning parameters as standard choices do not always lead to desirable behaviour in terms of operating characteristics. Such evaluations require comprehensive simulation studies and we have highlighted in this article important aspects to consider. 

 \begin{table}[h!]
\caption{Summary of recommendations for the robust component with normal distribution and unit-information or t-distributed with scale one and three degrees of freedom, for one-arm and hybrid-control trials in hypothesis testing and point estimation framework. }
\small
\resizebox{\textwidth}{!}{%
\begin{tabularx}{\textwidth}{|l|p{0.15\textwidth}|X|}
\hline
Aim&  Requirements &Recommendations \\
\hline
 \multirow{2}{3cm}[-10em]{Both hypothesis testing and point estimation} & \vspace{6em} Cap in TIE rate and RMSE needed & \begin{itemize} [topsep=-0.7cm,leftmargin=0.3cm]
    \item 	Set $\mu_{\text{robust}}=\bar{y}$ (or $\bar{y}_c$ in the hybrid control setting) \begin{itemize} [topsep=-0.1cm,leftmargin=0.3cm]
        \item Prior is data-dependent, though no problematic behaviour encountered in our simulations.
        \item more instances of bimodality compared to $\mu_{\text{robust}}=\bar{y}_{\text{ext}}$.
    \end{itemize}
    \item Set $\mu_{\text{robust}}=\bar{y}_{\text{ext}}$ but adopt a heavy-tailed distribution e.g. Student-t \begin{itemize} [topsep=-0.1cm,leftmargin=0.3cm]
        \item Prior is fixed.
        \item Requires MCMC or approximation by mixture of normals (need to choose number of components).
        \item Need to choose scale and degrees of freedom.
        \item Less instances of bimodality compared to $\mu_{\text{robust}}=\bar{y}$.
    \end{itemize}
\end{itemize} \\ \cline{2-3}
     & \vspace{0.15em} No cap in TIE rate and RMSE needed & \begin{itemize} [topsep=-0.6cm,leftmargin=0.3cm]
    \item Set $\mu_{\text{robust}}=\bar{y}_{\text{ext}}$ with a normal distribution (TIE rate increases to one, RMSE increases unbounded)
\end{itemize}  \\ \cline{1-3}
                 \multirow{2}{3cm}[-12em]{Only hypothesis testing} & \vspace{10em} Cap in TIE rate needed & \begin{itemize} [topsep=-0.6cm,leftmargin=0.3cm]
    \item 	Set $\mu_{\text{robust}}=\bar{y}$ (or $\bar{y}_c$ in the hybrid control setting)\begin{itemize} [topsep=-0.1cm,leftmargin=0.3cm]
        \item Prior is data-dependent, though no problematic behaviour encountered in our simulations.
        \item more instances of bimodality compared to $\mu_{\text{robust}}=\bar{y}_{\text{ext}}$.
    \end{itemize}
    \item Set $\mu_{\text{robust}}=\bar{y}_{\text{ext}}$ but adopt a heavy-tailed distribution e.g. Student-t \begin{itemize} [topsep=-0.1cm,leftmargin=0.3cm]
        \item Prior is fixed.
        \item Requires MCMC or approximation by mixture of normals (need to choose number of components).
        \item Need to choose scale and degrees of freedom.
        \item Less instances of bimodality compared to $\mu_{\text{robust}}=\bar{y}$.
    \end{itemize}
    \item Set $\mu_{\text{robust}}=\theta_0$ \newline Note: this is not an option for the hybrid-control setting since the border of the null hypothesis is a line, i.e. $\theta_t=\theta_c.$ \begin{itemize} [topsep=-0.1cm,leftmargin=0.3cm]
        \item Prior is fixed.
    \end{itemize}
\end{itemize}  \\ \cline{2-3}
                                   & \vspace{0.5em} No cap in TIE rate needed & \begin{itemize} [topsep=-0.6cm,leftmargin=0.3cm]
    \item Set $\mu_{\text{robust}}=\bar{y}_{\text{ext}}$ with a normal distribution (TIE rate increases to one)
 
\end{itemize} \\
   \hline
\end{tabularx}
}
\label{Table:recommendations}
\end{table}
\newpage
   \bibliographystyle{apalike}%
\bibliography{main}%
\newpage
\section*{Appendix}



\appendix
\renewcommand\thefigure{A.\arabic{figure}}    
\setcounter{figure}{0}   
\renewcommand\thetable{A.\arabic{table}}    
\setcounter{table}{0}  

\begin{figure}[h!] 
\includegraphics[width=\textwidth,height=\textheight,keepaspectratio]{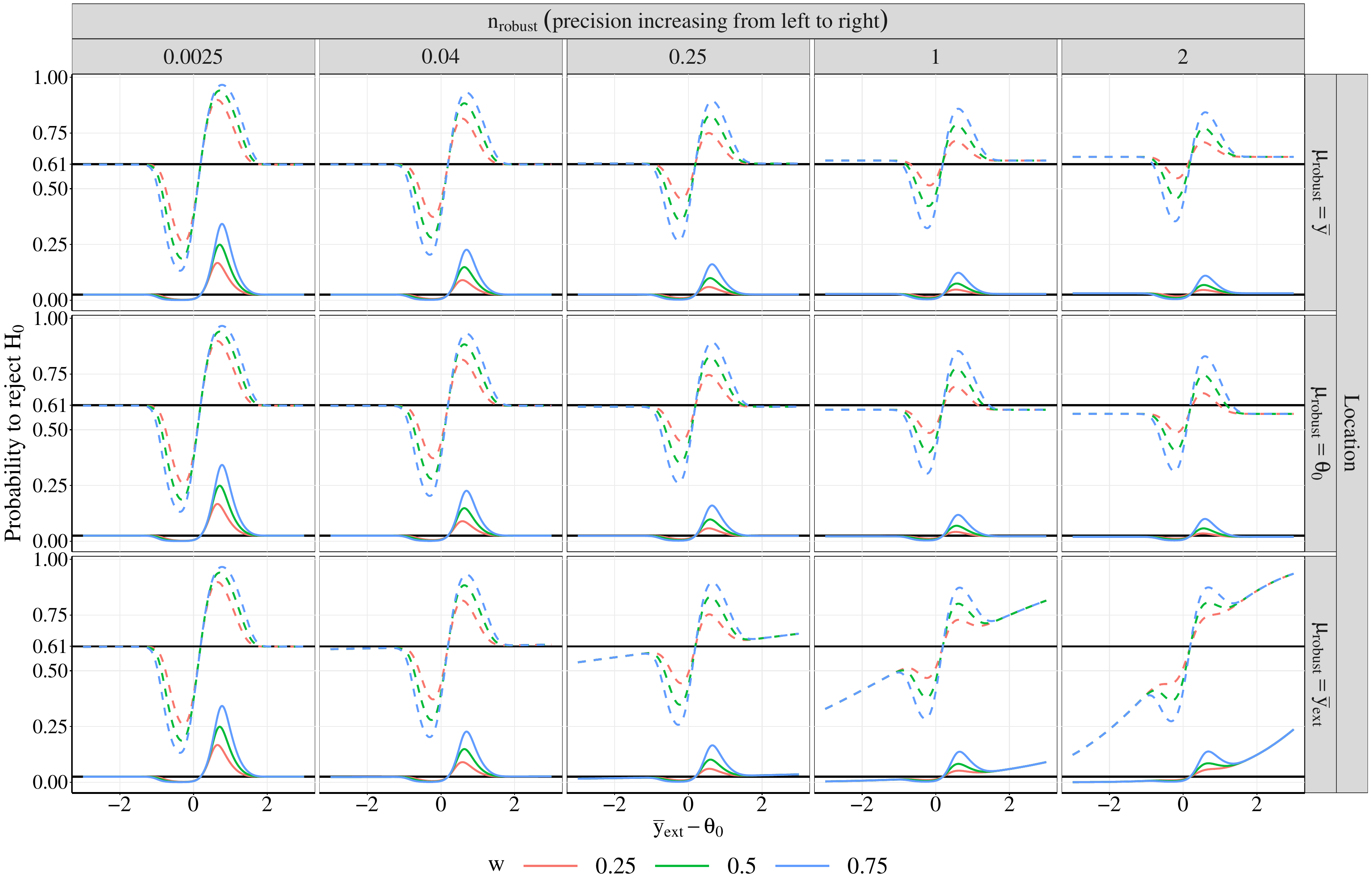}
	\caption{Power (dashed lines) and TIE rate (solid lines) for the one-arm trial with different locations and dispersions of robust component as well as different prior weights $(0.25, 0.5, 0.75)$ and $n_{\text{current}}=20$ and $n_{\text{ext}}=15$. The column $n_{\text{robust}}=1$ represents the unit-information.}
 \label{power_APPENDIX}
\end{figure}
\newpage
\begin{figure}[h!]
\includegraphics[width=\textwidth,height=\textheight,keepaspectratio]{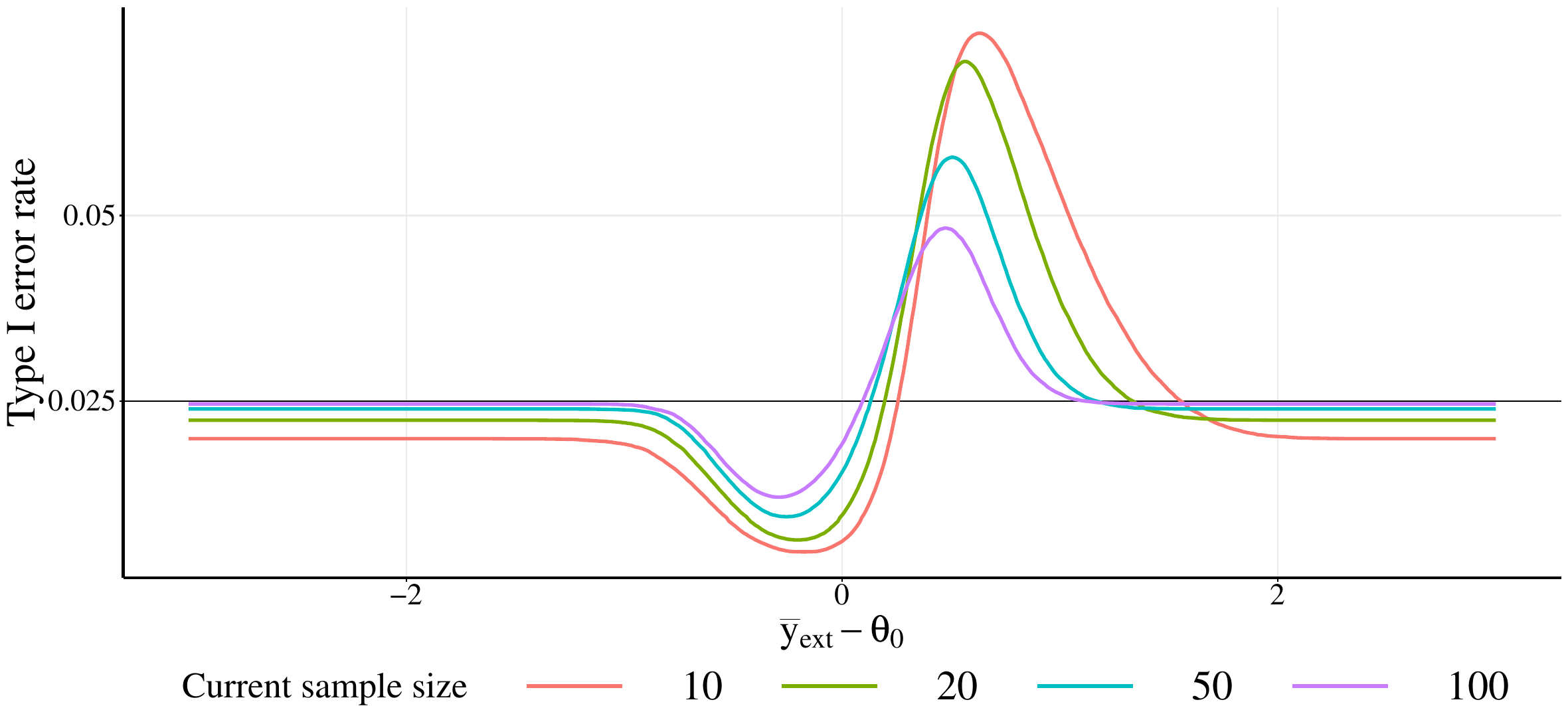}
	\caption{TIE rate for the one-arm trial with the robust component of the mixture as a unit-information prior. The unit-information prior will have non-negligible impact on the posterior in case of small current sample size. Here $n_{\text{ext}}=15$, prior weight $w=0.5$ and $\mu_{\text{robust}}=\theta_0$. }
 \label{TIE_unit_ncurr}
\end{figure}
\newpage
\begin{figure}[h!]
\includegraphics[width=\textwidth,height=\textheight,keepaspectratio]{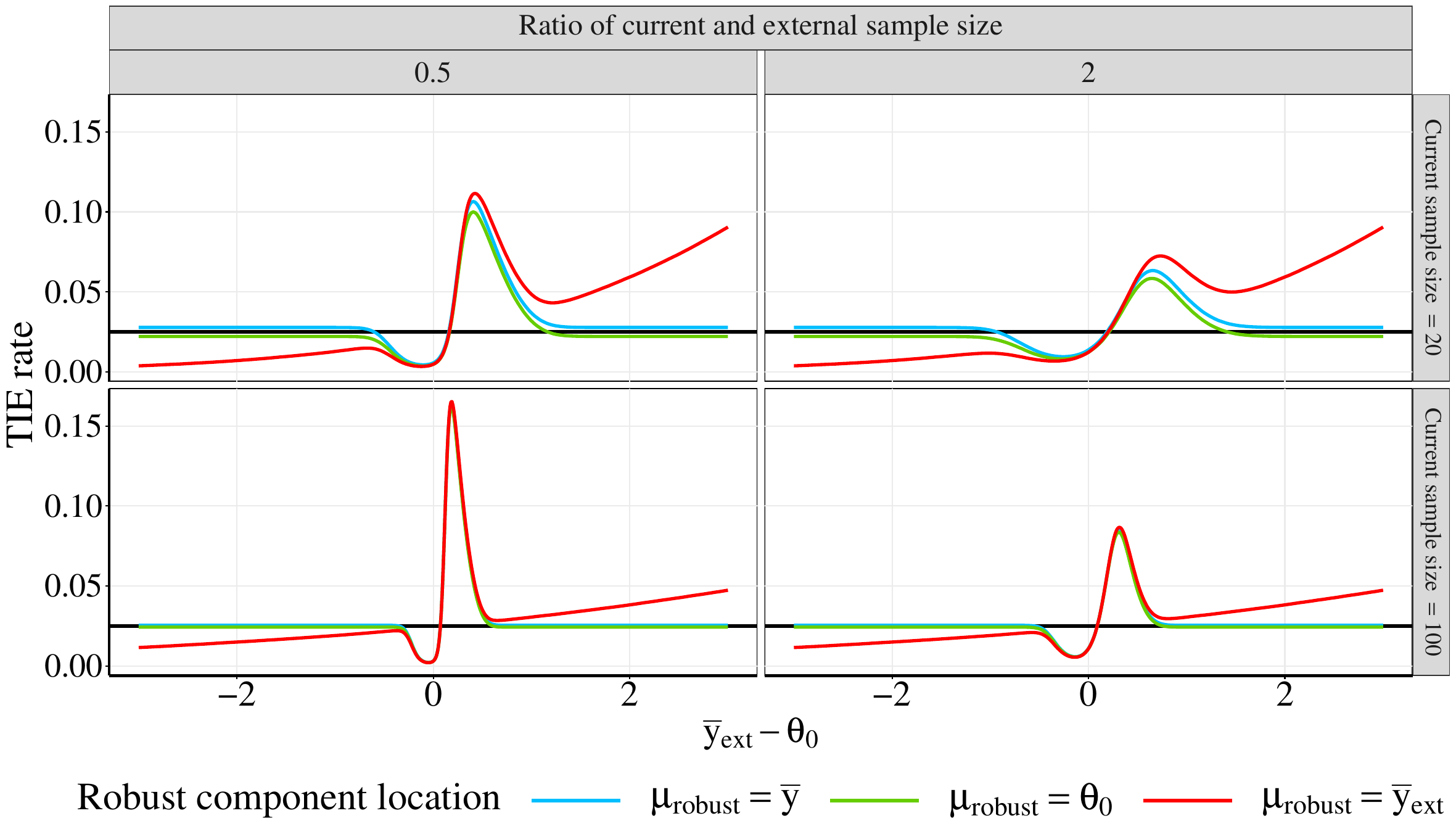}
	\caption{TIE rate for the one-arm trial with the robust component of the mixture as a unit-information prior but with different ratios of current to external sample size. Here prior weight $w=0.5$. }
 \label{TIE_rate_different_sample_sizes}
\end{figure}
\newpage
\begin{figure}[h!]
\includegraphics[width=\textwidth,height=\textheight,keepaspectratio]{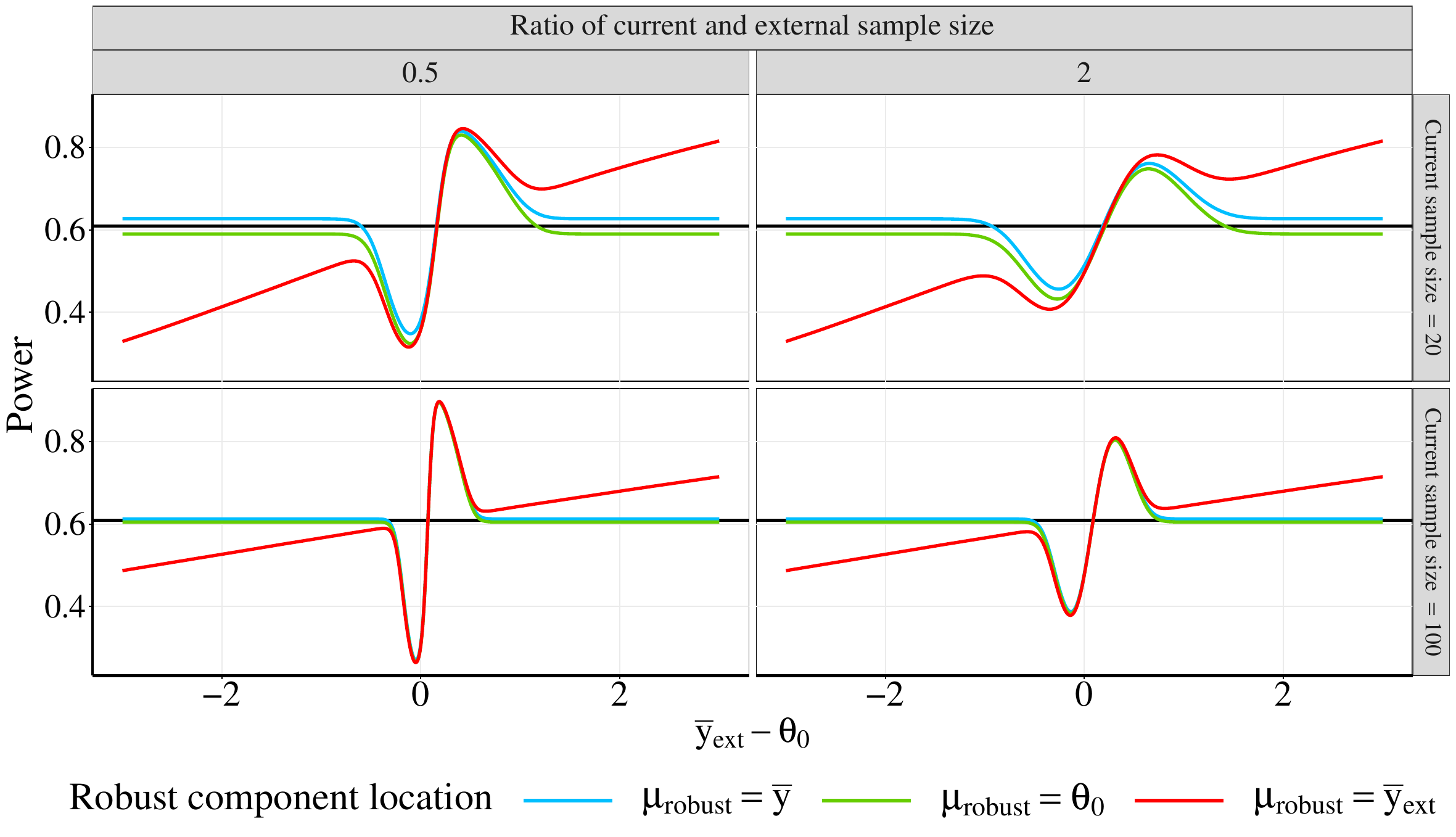}
	\caption{Power for the one-arm trial with the robust component of the mixture as a unit-information prior but with different ratios of current to external sample size. Here prior weight $w=0.5$. }
 \label{Power_different_sample_sizes}
\end{figure}
\newpage
\begin{figure}[h!]   
\includegraphics[width=\textwidth,height=\textheight,keepaspectratio]{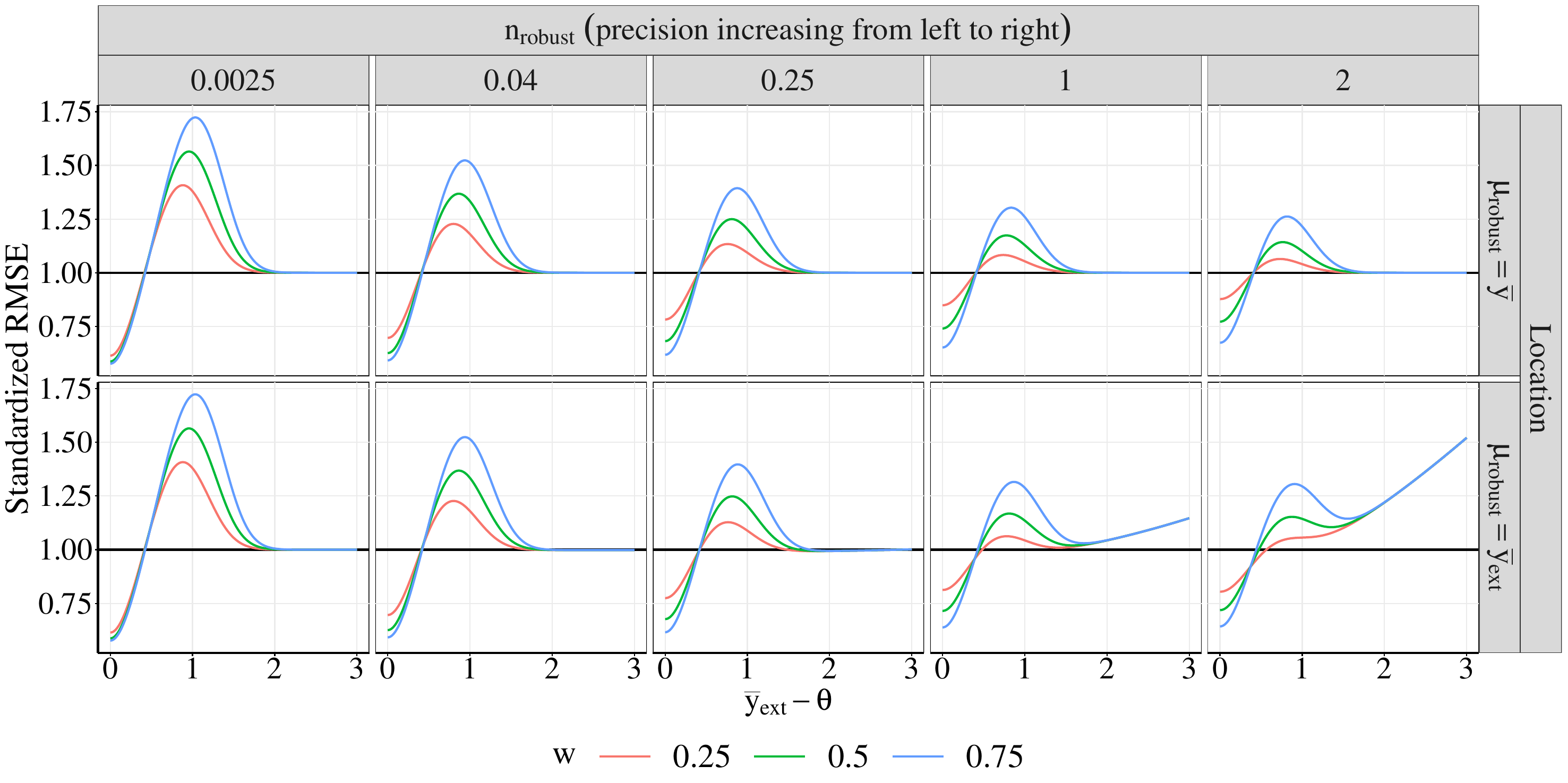}
	\caption{Standardized RMSE with different locations and dispersions of the robust component of the mixture prior as well as different prior weights for the informative component $(0.25, 0.5, 0.75)$  and $n_{\text{current}}=20$ and $n_{\text{ext}}=15$. The column $n_{\text{robust}}=1$ represents the unit-information.}
 \label{MSE_APPENDIX}
\end{figure}
\newpage
\begin{figure}[h!]
\includegraphics[width=\textwidth,height=\textheight,keepaspectratio]{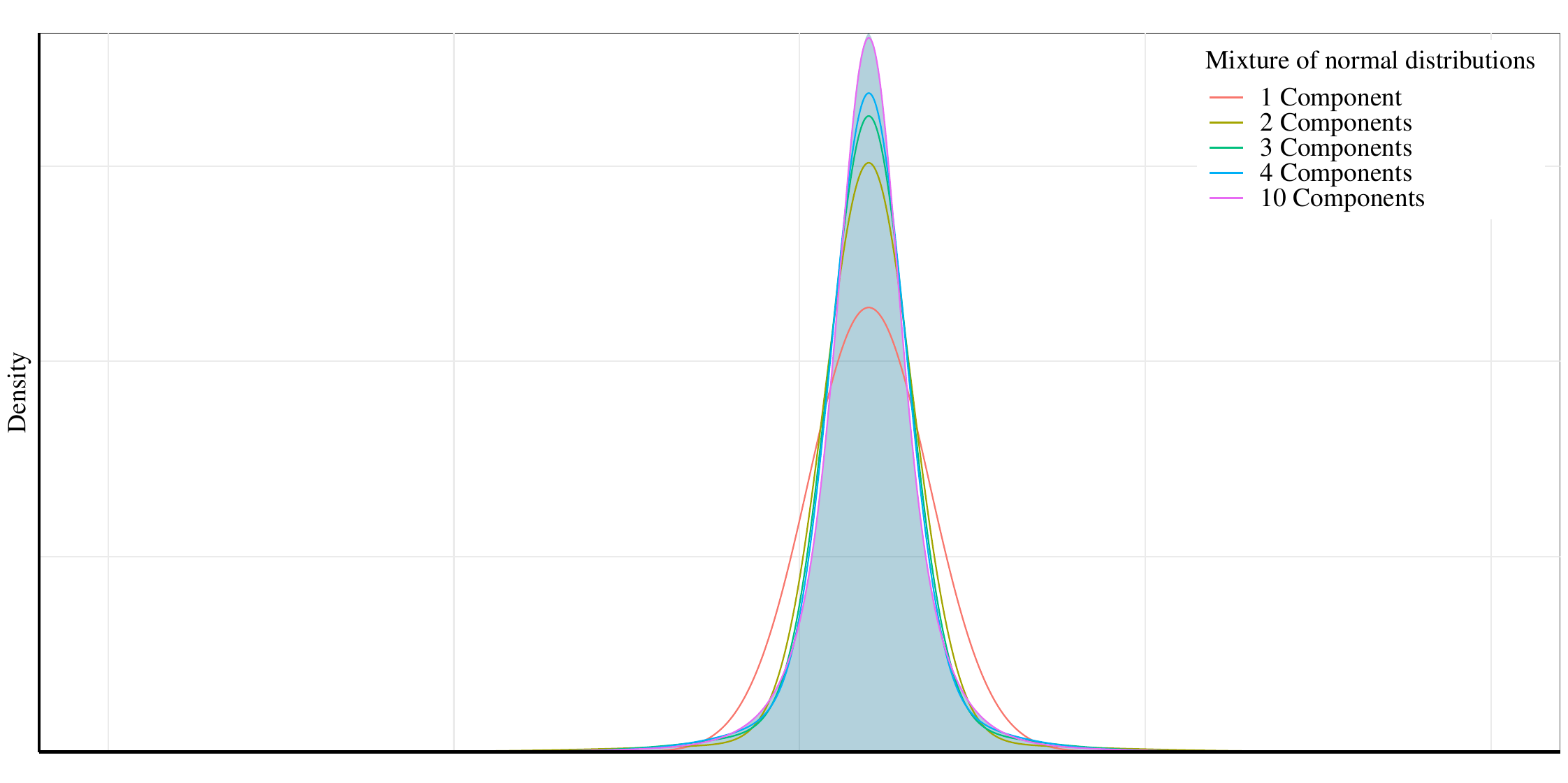}
	\caption{Approximation of a t-distribution by a mixture of normal distributions with different number of components. The blue mass represent the density of the t-distribution with $\text{scale}=1$ and $3$ degrees of freedom. }
 \label{t_approx_APPENDIX}
\end{figure}
\newpage
\begin{figure}[h!]
\includegraphics[width=0.9\textwidth,height=0.9\textheight,keepaspectratio]{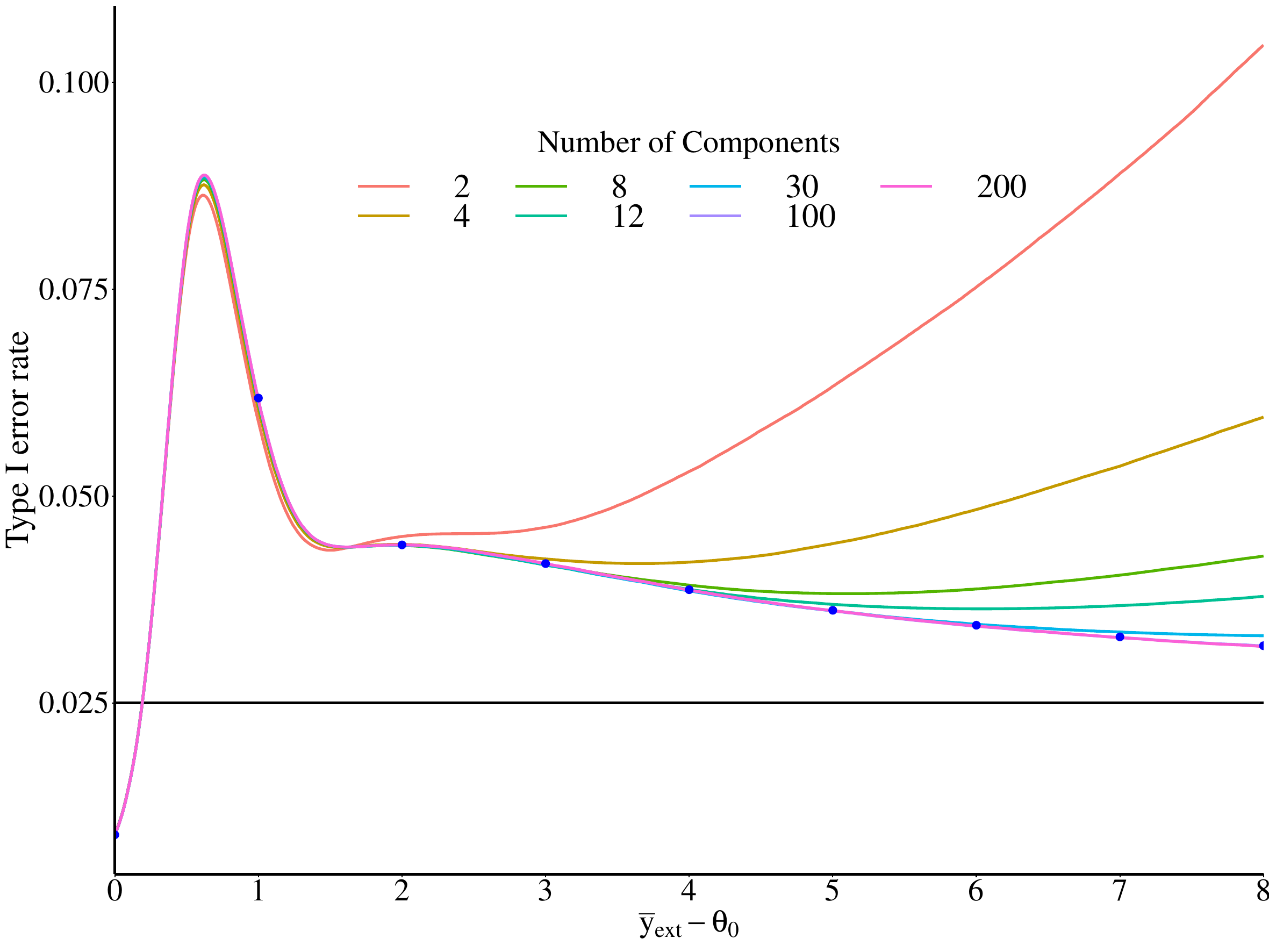}
	\caption{TIE rate for the one-arm trial setting with the robust component as a t-distribution approximated by a mixture of normal distributions with different number of components indicated by the different lines. The blue dot is the result using the exact t-distribution, based on MCMC (shown for a coarse grid). Here $\mu_{\text{robust}}=\bar{y}_{\text{ext}}$, $n_{\text{current}}=20$, $n_{\text{ext}}=15$ and $w=0.5$. Results are shown for extended prior-data conflict in the alternative hypothesis direction. The t-distribution has scale=1 and 3 degrees of freedom.}
 \label{number_of_components_onearm}
\end{figure}

\newpage
\begin{figure}[h!]
 \includegraphics[width=\textwidth,height=\textheight,keepaspectratio]{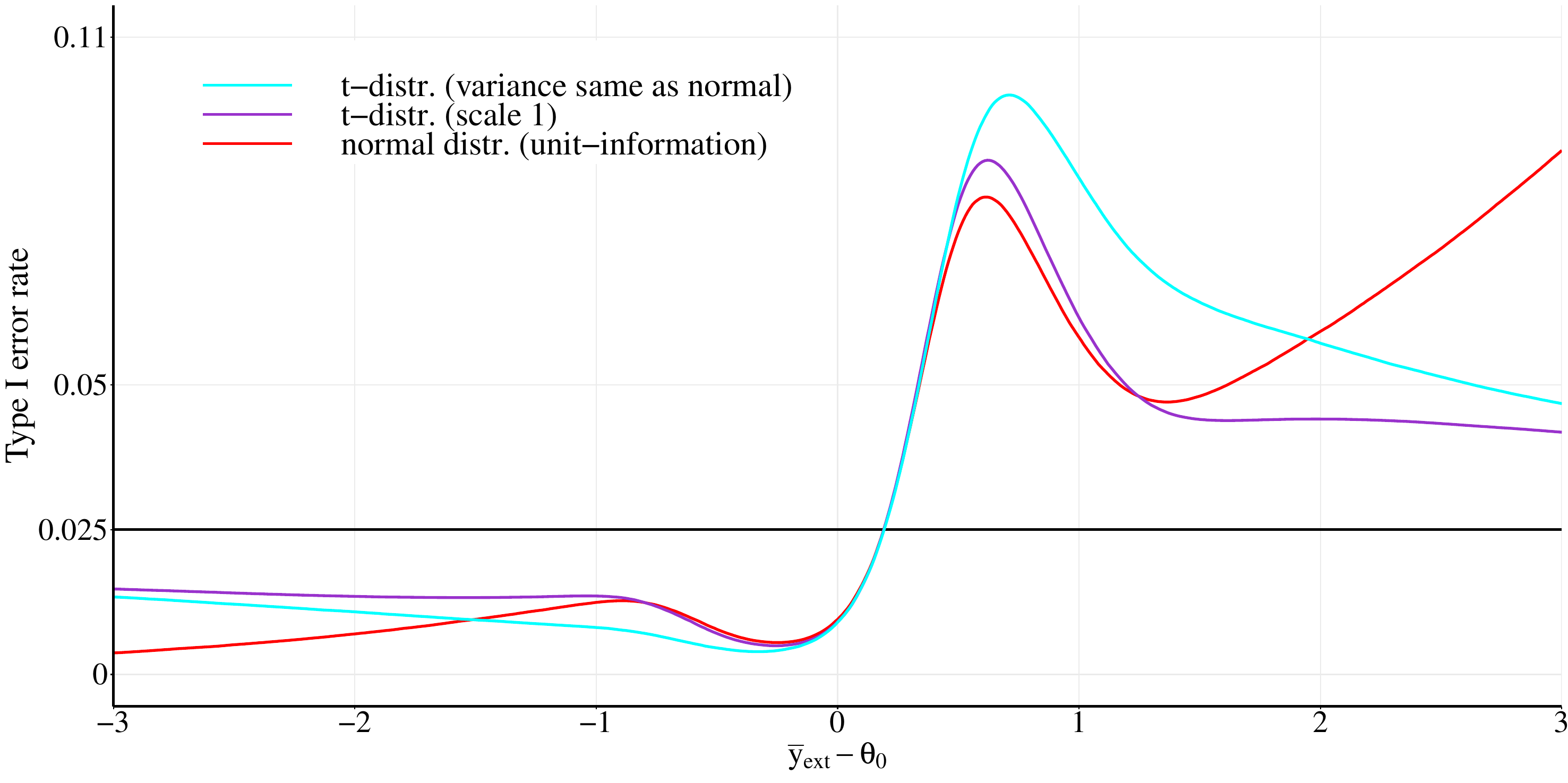}
	\caption{ TIE rate for the one-arm trial where the robust component is a unit-information prior or t-distributed with different scales but 3 degrees of freedom. Here $n_{\text{current}}=20$, $n_{\text{ext}}=15$, prior weight $w=0.5$ and $\mu_{\text{robust}}=\bar{y}_{\text{ext}}$.}
 \label{tdist_APPENDIX}
\end{figure}
\newpage
\begin{figure}[h!]
    \includegraphics[width=\textwidth,height=\textheight,keepaspectratio]{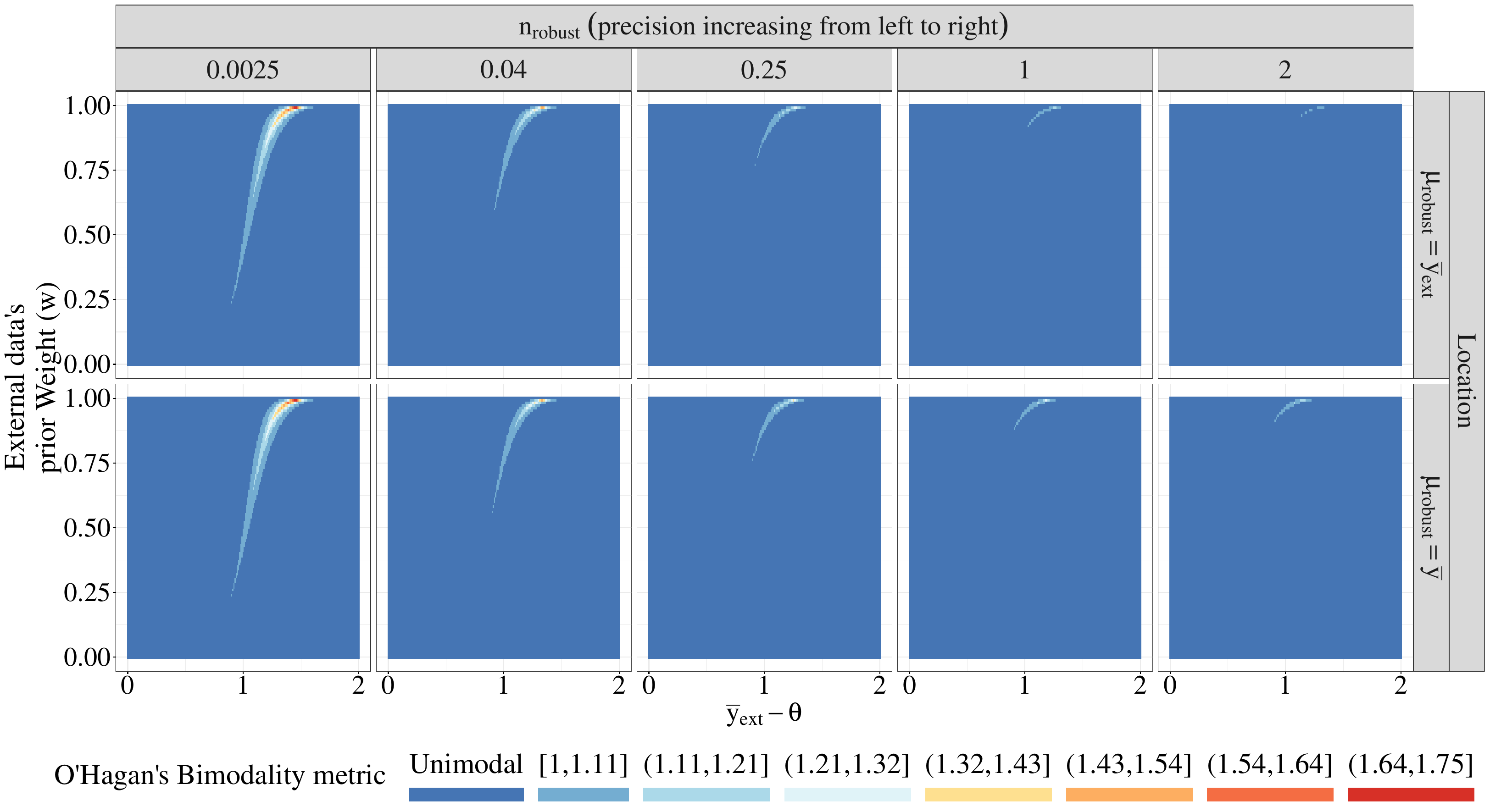}
	\caption{Bimodality: OBM metric of the posterior distribution for different location and dispersion for the robust component of the mixture prior and different prior weights assigned to the informative component. The column $n_{\text{robust}}=1$ represents the unit-information. Again, $n_{\text{current}}=20$ and $n_{\text{ext}}=15$. We note that in the evaluation of bimodality, we consider $\bar{y}=\theta$.}
 \label{bimodality_appendix}
\end{figure}
\newpage

\newpage
\begin{figure}[h!]
    \includegraphics[width=\textwidth,height=\textheight,keepaspectratio]{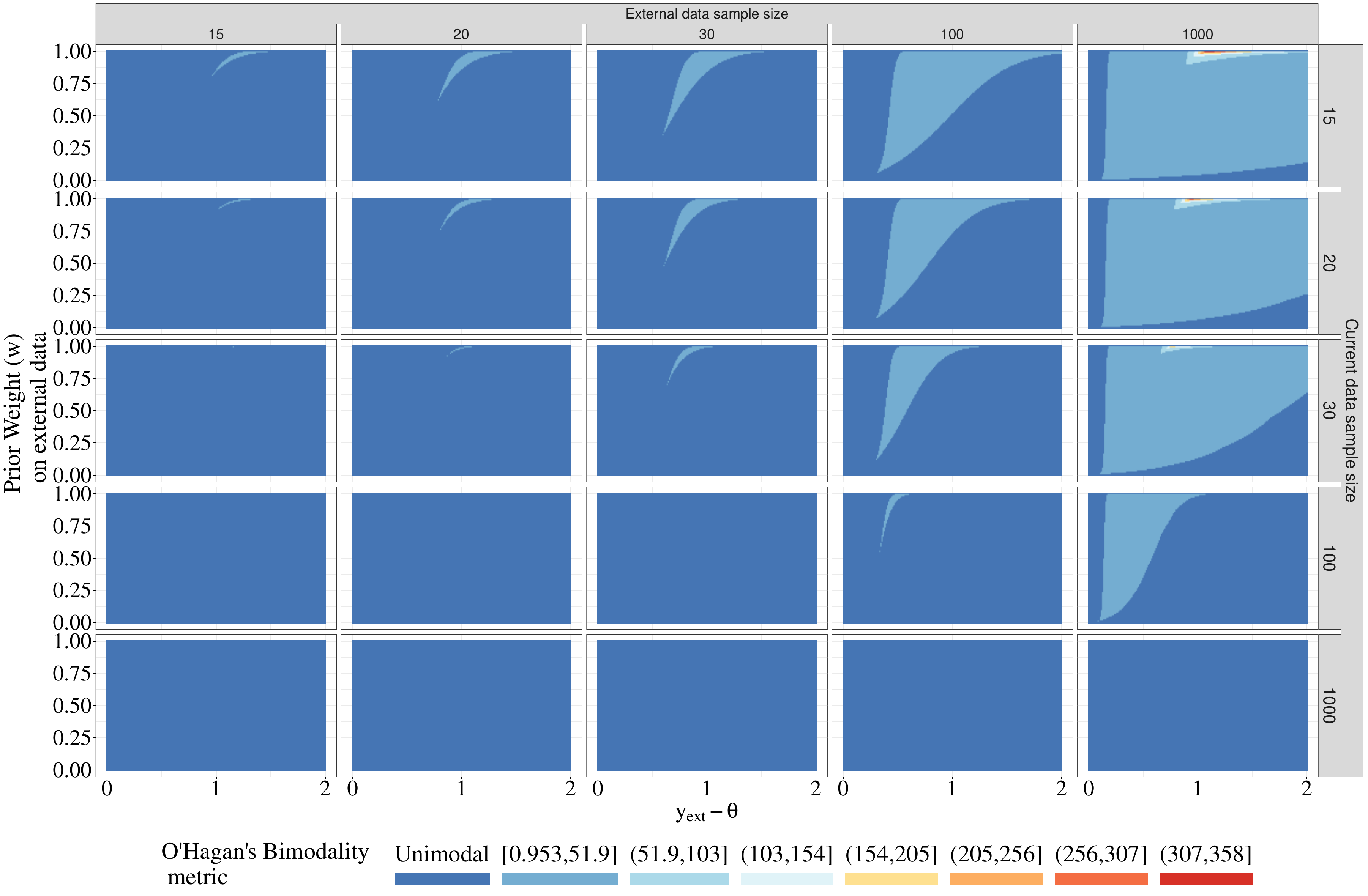}
	\caption{Bimodality: OBM metric of the posterior distribution for a unit-information robust component with different current and external sample sizes. On the y-axis is a grid of prior weights assigned to the informative component while prior-data conflict is shown on the x-axis. Here $\mu_{\text{robust}}=\bar{y}_{\text{ext}}$. We note that in the evaluation of bimodality, we consider $\bar{y}=\theta$.}
 \label{bimodality_varying_n_appendix}
\end{figure}

\newpage
\begin{figure}[h!]
   \includegraphics[width=\textwidth,height=\textheight,keepaspectratio]{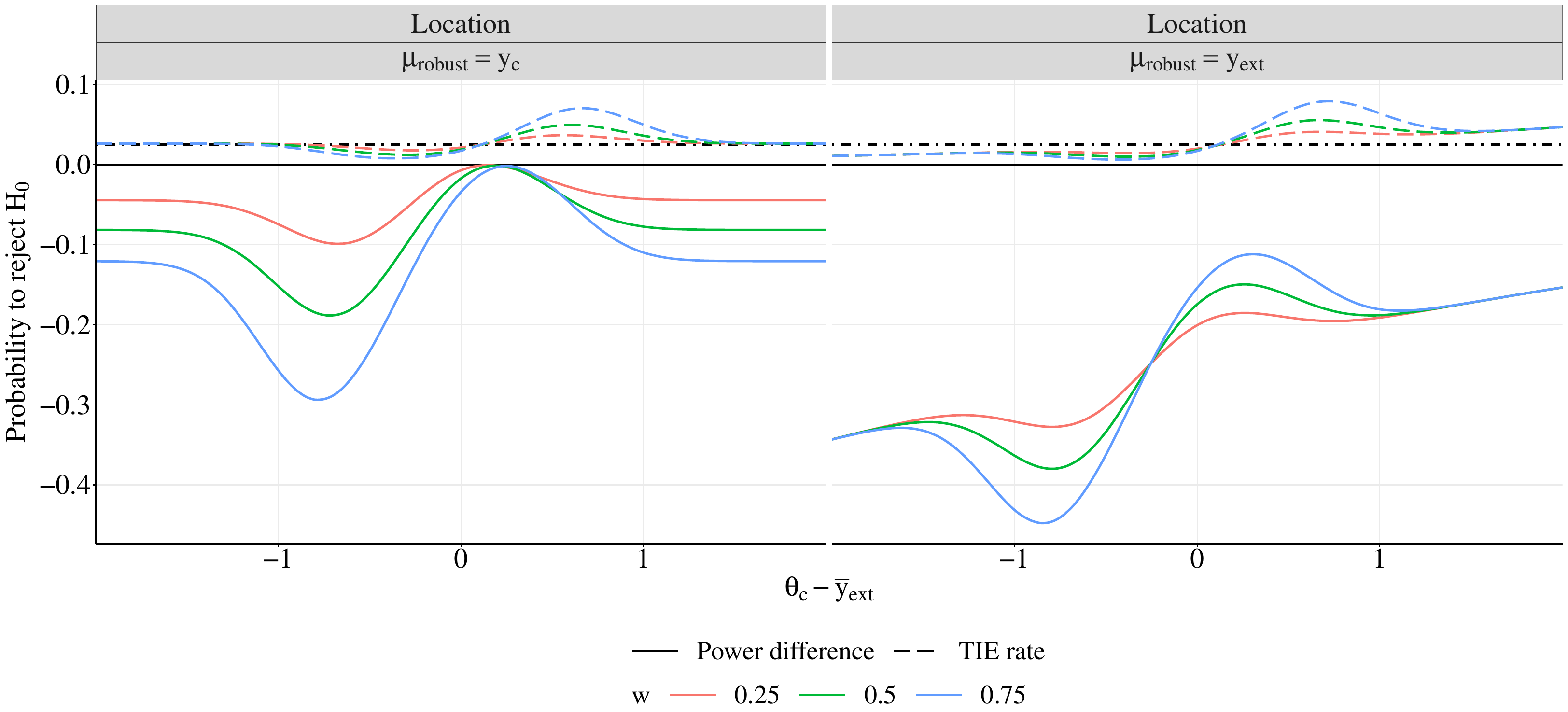}
	\caption{Hybrid control trial with the robust component of the mixture prior as a unit-information prior. Power difference of the test with borrowing and calibrated to borrowing as well as the TIE rate are shown for the two robust component locations and different prior weights. Here, $n_t=n_c=20$ and $n_{\text{ext}}=15$. The frequentist power of $0.75$ is obtained at $\theta_t-\theta_c=0.83$ and $\alpha=0.025$.  }
 \label{power_diff:2arm}
\end{figure}

\newpage
\begin{figure}[h!]
   \includegraphics[width=\textwidth,height=\textheight,keepaspectratio]{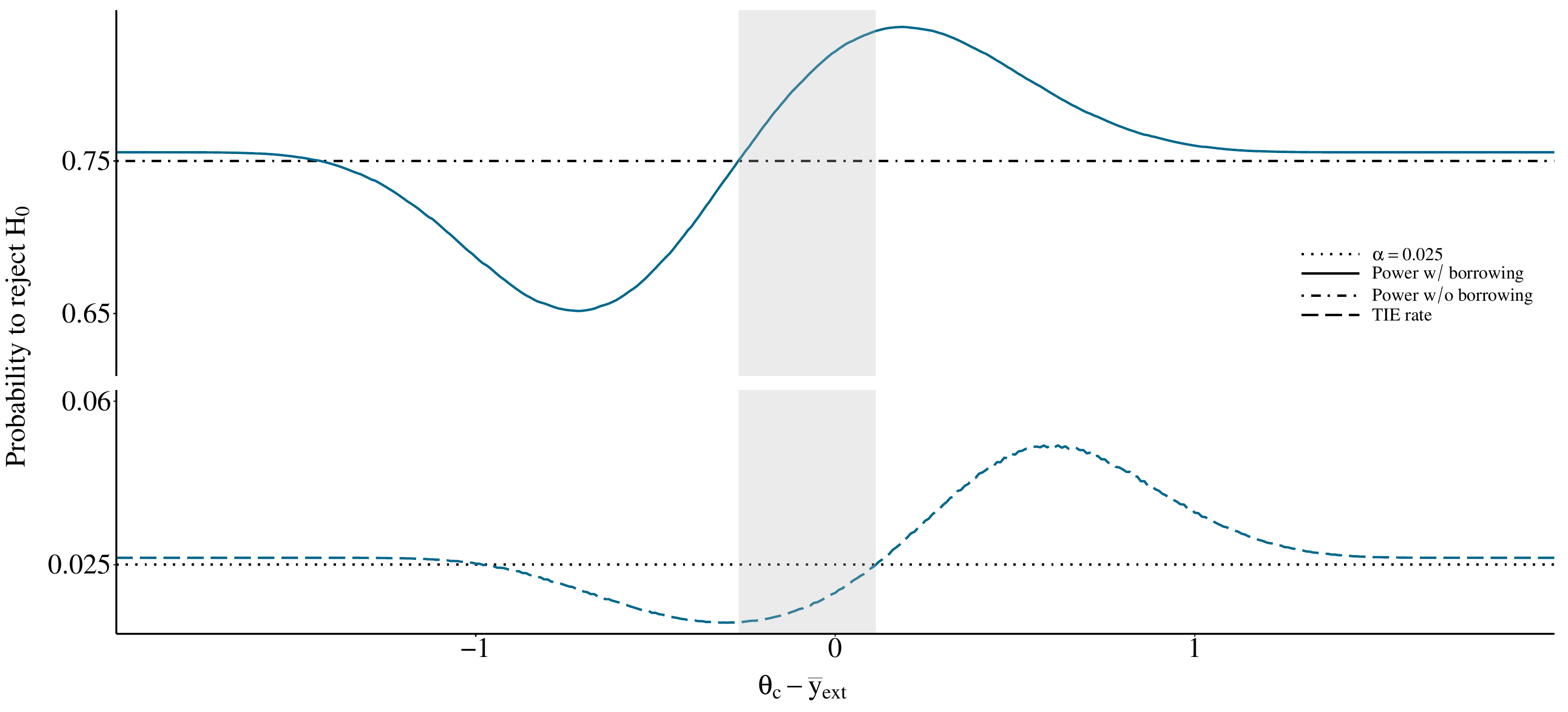}
	\caption{An illustration of the sweet spot is shown in lightgray for $w=0.5$ and unit-information robust component with $\mu_{\text{robust}}=\bar{y}_c$. The frequentist power of $0.75$ is obtained at $\theta_t-\theta_c=0.83$ for $n_t=n_c=20$ and $\sigma^2=1$ at $\alpha=0.025$.   }
 \label{sweetspot_illustration}
\end{figure}

\newpage
\begin{figure}[h!]
   \includegraphics[width=\textwidth,height=\textheight,keepaspectratio]{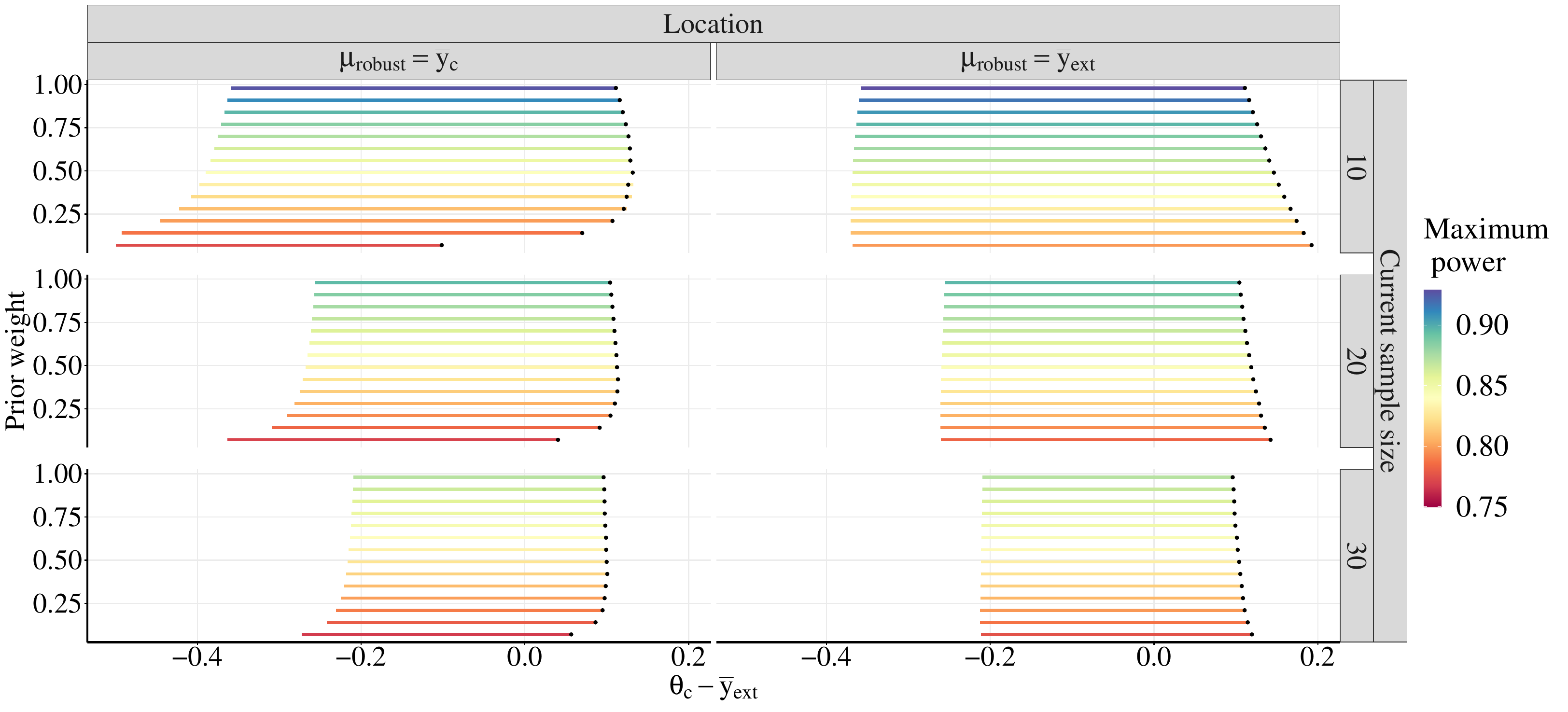}
	\caption{In the hybrid control trial, the size of the sweet spot is shown for the different locations for the unit-information robust component, different prior weights assigned to the informative component and different sample sizes of the current trial. The color indicates the maximum power gain in the sweet spot while the black dot shows where in this spot the maximum power gain is observed. For contrast, power of the one-sided two-sample test without borrowing is evaluated at different $\theta_t-\theta_c=\theta_1 \in H_1$ for the different sample sizes so that the same power of $0.75$ is obtained with $\sigma^2=1$ at $\alpha=0.025$. }
 \label{a:sweetspot}
\end{figure}
\newpage
\begin{figure}[h!]
\centering\includegraphics[width=\textwidth,height=\textheight,keepaspectratio]{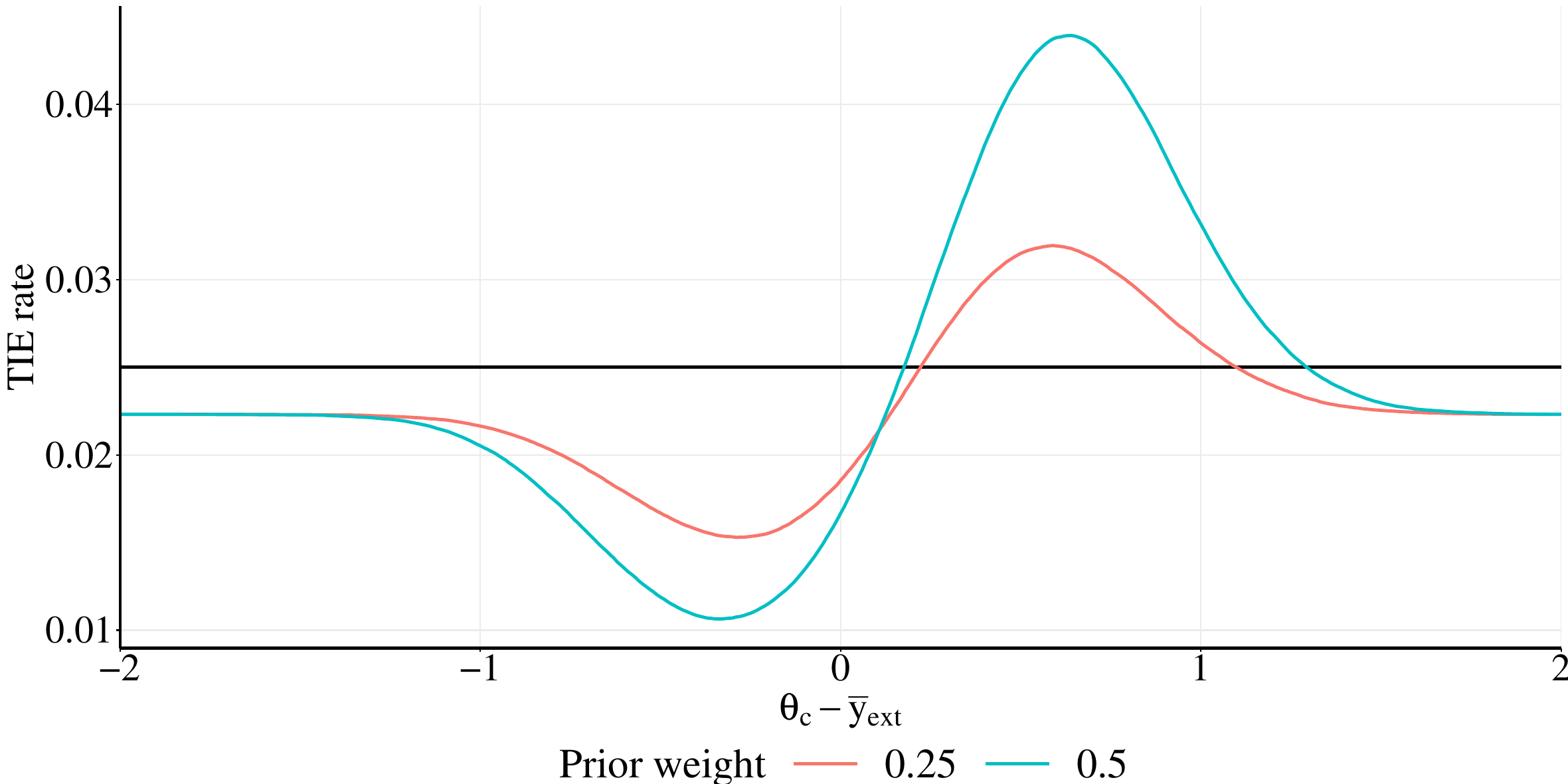}
	\caption{TIE rate is shown when the unit-information robust component located at $\mu_{\text{robust}}=\bar{y}_{\text{ext}}$ is also used as prior for the treatment arm. Here $n_t=n_c=20, n_{\text{ext}}=15$. Different prior weights are investigated represented by the different colors.}
 \label{unit:treatarm}
\end{figure}
\newpage
\begin{figure}[h!]
   \includegraphics[width=\textwidth,height=\textheight,keepaspectratio]{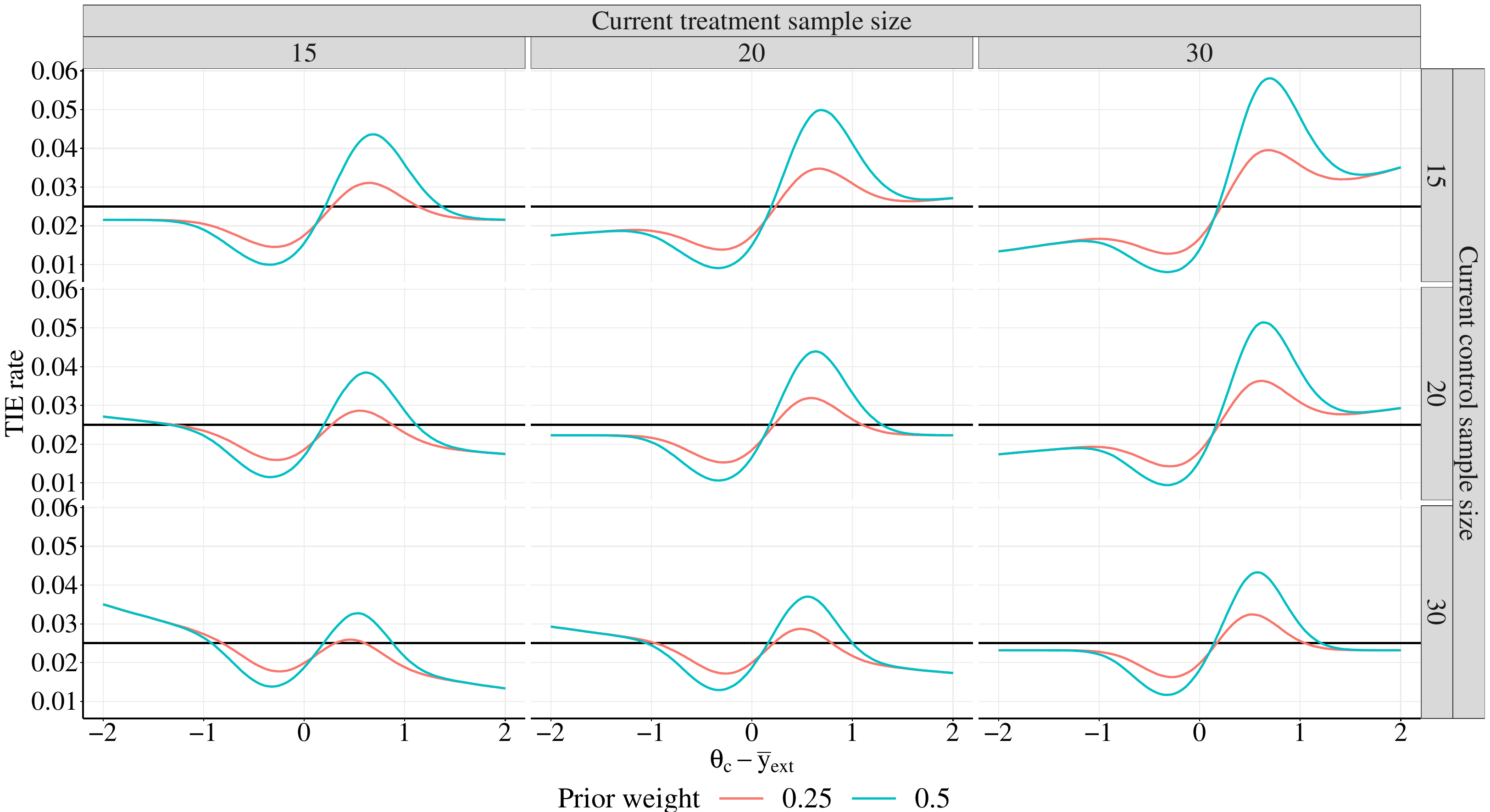}
	\caption{TIE rate is shown when the unit-information robust component located at $\mu_{\text{robust}}=\bar{y}_{\text{ext}}$ is also used as prior for the treatment arm. Here $n_{\text{ext}}=15$ while $n_t$ and $n_c$ varies. Different prior weights are investigated represented by the different colors.}
 \label{unit_treatarm_unbal}
\end{figure}
\newpage
\begin{figure}[ht!]
   \centering
  \includegraphics[width=\textwidth,height=\textheight,keepaspectratio]{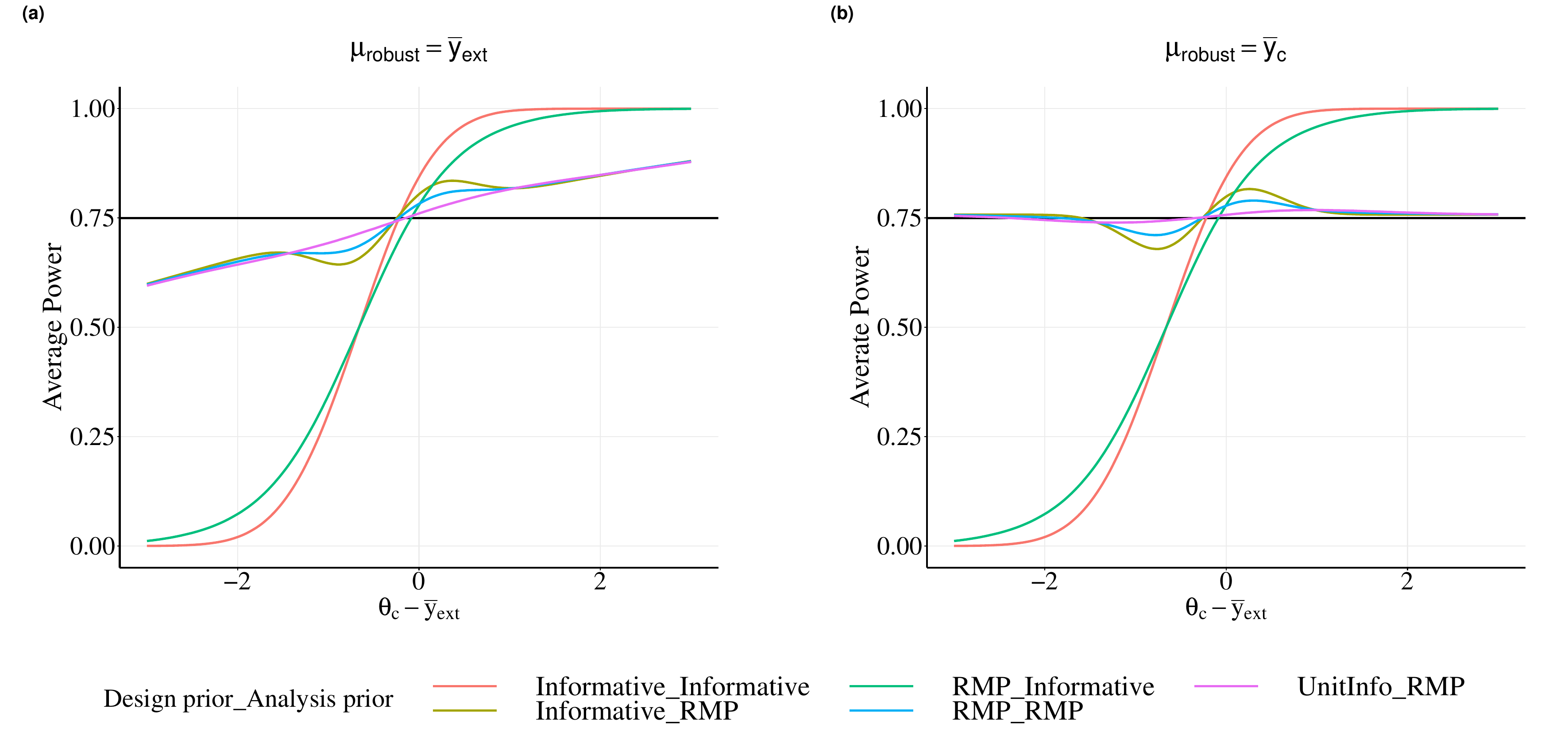}
	\caption{Average power for the hybrid control trial setting is shown for different design and analysis priors. The analysis prior is additionally shifted to reflect a range of prior-data conflict. (a) and (b) represent the two different robust component locations for the robust mixture analysis prior, i.e. $\mu_{\text{robust}}=\bar{y}_{\text{ext}}$ and  $\mu_{\text{robust}}=\bar{y}_{\text{c}}$, respectively. The legend shows the respective priors separated by underscore (i.e. design prior\_analysis prior). Where the robust mixture prior (RMP) is shown, a weight of $0.5$ was used. Here $n_t=n_c=20$ and $n_{\text{ext}}=15$.} 
 \label{average_power:2arm}
\end{figure}

\end{document}